\documentclass[12pt]{article}

\usepackage[margin=1in]{geometry}
\usepackage[caption = false]{subfig}
\usepackage{amsmath}
\usepackage{amsfonts}
\usepackage{amssymb}
\usepackage{graphicx}
\usepackage{bbm}
\usepackage{enumerate}
\usepackage{hyperref}
\usepackage{mathtools}
\usepackage{color}
\usepackage[table]{colortbl}
\usepackage[dvipsname, table]{xcolor}
\usepackage{chngcntr}
\usepackage{hhline}
\usepackage{multirow}
\usepackage{makecell}
\usepackage{comment}
\usepackage{algorithm2e}

\usepackage{etoolbox}
\makeatletter
\patchcmd{\@makecaption}
  {\parbox}
  {\advance\@tempdima-\fontdimen2} 
  {}{}
\makeatother  

\newtheorem{thm}{Theorem}[section]
\newtheorem{prop}{Proposition}[section]

\newtheorem{assumption}{Assumption}

\newenvironment{proof}{\paragraph{Proof:}}{\hfill$\square$}

\definecolor{orange}{rgb}{255, 165, 0}
\definecolor{red}{rgb}{255, 0, 0}
\definecolor{yellow}{rgb}{255, 255, 0}

\newcommand{\E}{\mathbb{E}}
\newcommand{\R}{\mathbb{R}}

\newcommand{\boldG}{\boldsymbol{G}}

\newcommand{\boldI}{\boldsymbol{I}}

\newcommand{\boldV}{\boldsymbol{V}}
\newcommand{\boldW}{\boldsymbol{W}}
\newcommand{\boldX}{\boldsymbol{X}}

\newcommand{\bolda}{\boldsymbol{a}}

\newcommand{\boldt}{\boldsymbol{t}}

\newcommand{\boldx}{\boldsymbol{x}}
\newcommand{\boldy}{\boldsymbol{y}}

\newcommand{\indicator}{\mathbbm{1}}

\newcommand{\bbeta}{\boldsymbol{\beta}}

\newcommand{\bdeta}{\boldsymbol{\eta}}

\newcommand{\bepsilon}{\boldsymbol{\varepsilon}}

\newcommand{\btheta}{\boldsymbol{\theta}}
\newcommand{\bmu}{\boldsymbol{\mu}}

\newcommand{\bSigma}{\boldsymbol{\Sigma}}

\newcommand{\boldzero}{\mathbf{0}}

\newcommand{\prob}{\mathbb{P}}

\DeclareMathOperator*{\argmin}{arg\,min}
\DeclareMathOperator*{\argmax}{arg\,max}

\begin{document}

\title{ENNS: Variable Selection, Regression, Classification and  Deep Neural Network for High-Dimensional Data}

\author{Kaixu Yang
and 
Tapabrata Maiti\\
yangkaix@msu.edu\\
 maiti@msu.edu\\
619 Red Cedar Rd.\\
Department of Statistics and Probability\\
Michigan State University\\
East Lansing, MI, 48824}

\maketitle

\begin{abstract}%
High-dimensional, low sample-size (HDLSS) data problems have been a topic of immense importance for the last couple of decades. There is a vast literature that proposed a wide variety of approaches to deal with this situation, among which variable selection was a compelling idea. On the other hand, a deep neural network has been used to model complicated relationships and interactions among responses and features, which is hard to capture using a  linear or an additive model. In this paper, we discuss the current status of variable selection techniques with the neural network models. We show that the stage-wise algorithm with neural network suffers from disadvantages such as the variables entering into the model later may not be consistent. We then propose an ensemble method to achieve better variable selection and prove that it has probability tending to zero that a false variable is selected. Then, we discuss additional regularization to deal with over-fitting and make better regression and classification. We study various statistical properties of our proposed method. Extensive simulations and real data examples are provided to support the theory and methodology.
\end{abstract}

\noindent%

\section{Introduction}
\label{section:introduction}

High-dimensional data modeling in statistics has been popular for decades, \cite{buhlmann2011statistics}. Consider a high-dimensional regression or a binary classification problem. Let $\boldx\in\R^p$ be the feature vector, and let $y\in\R$ for regression problem and $y\in\{0,1\}$ for classification problem be the response. Our goal is to build a parsimonious model based on the training sample $\{(\boldx_1,y_1),...,(\boldx_n,y_n)\}$. We have more features than the sample size, i.e., $p>n$. Moreover, many data have a complicated relationship among different variables, which is hard to capture through explicit modeling. Neural network modeling is one of the best ways of capturing complex relationships among variables with minimal mathematical assumptions. Thus, it is interesting to consider a neural network structure between $\boldx$ and $y$.

In general, high-dimensional model does not have consistent estimations since we have less number of observations than the number of variables without any assumptions. Two major approaches can be used to deal with the high-dimensionality. The first major approach is to assume that the feature space is sparse, i.e., only a small fraction of variables are effectively included in modeling with $y$. A model with only a fraction of the original features enjoys simplicity and interpretability. Sparse solutions can be obtained using soft-thresholding methods \cite{donoho1995noising} or regularization \cite{tibshirani1996regression, huang2010variable} or stage-wise algorithms \cite{efron2004least}. Regularization obtains a sparse solution by shrinking the unimportant features' coefficients to zero. The estimated coefficients are shrinkage estimators and thus have smaller variance \cite{copas1983regression}. However, regularization with multiple tuning parameters takes longer to run and maybe sensitive in tuning the parameters in practice. Stage-wise algorithms add variables one by one and stop at a preferred stopping time. 

The second dominant approach is projection-based. One finds a lower-dimensional representation of the original feature space. Linear projection methods include the PCA \cite{hotelling1933analysis} in the low dimensional case and some of its variants \cite{jolliffe2003modified, zou2006sparse} in the high-dimensional case. Kernel PCA \cite{scholkopf1998nonlinear} performs PCA on a reproducing kernel Hilbert space to achieve non-linearity. Manifold learning, \cite{lawrence2012unifying} embeds the original feature space to a low-dimensional manifold. Except for the manifold learning algorithms that reduce the original dimensionality to two or three dimension for visualization, a few manifolding learning including the multidimensional sacling (MDS) by \cite{torgerson1952multidimensional}, the local linear embedding (LLE) by \cite{roweis2000nonlinear}, and the Isomap by \cite{tenenbaum2000global} are applied to low-dimensional dimensionality reduction. Applications of the manifold learning algorithms in the high-dimensional set up is studied for specific fields, but a general framework is not available. The current manifold learning algorithms focus more on data visualization, which reduces the dimensionality to two or three, see, for example, \cite{wang2019capacity}. Another popular dimension reduction technique is the auto-encoder \cite{kramer1991nonlinear}, which uses a neural network to encode the feature space and decode the representation to be as close to the original feature space as possible. All of the above methods are unsupervised, and the lower-dimensional representation is no longer preserves any of the original features. Therefore, we lose interpretability by adopting this approach.  Thus, these applications are not useful in building regression models.

On the other hand, neural networks have been utilized to model complicated relationship since  1940s \cite{kleene1951representation}, and gained much more attention since the great improvements in computer hardware  in this century. Specifically, \cite{oh2004gpu} showed that the computation of neural networks could be greatly improved by GPU (graphics computing unit)  acceleration than purely running on CPU (central processing unit), this makes it easy to train deeper network structures. Nowadays, variants of neural networks are being applied world-wide, including the convolutional neural network (CNN), recurrent neural network (RNN), residual network (ResNet), etc. In theory, a neural network works in representing complicated relationships, mainly lies on the universal approximation theorem \cite{cybenko1989approximation, barron1993universal, anthony2009neural, siegel2019approximation}. The theorem states that a shallow neural network (neural network with one hidden layer) is able to approximate any continuous function with an arbitrarily small error given a large number of hidden nodes, under mild assumptions. In practice, to achieve this good approximation, usually, a massive set of training data is needed, since the number of parameters in a neural network is much more than that in other conventional statistical models. Moreover, the non-convexity of a neural network structure makes it impossible to obtain a global optimum. Luckily, a local optimum of the neural network provides a good approximation in a reasonable practical sense.

Deep Neural Network (DNN) is a neural network with a deep structure of hidden layers, which has better performance than the shallow neural network (neural network with only one hidden layer) in many aspects in a broad field of applications such as pattern recognition, speech recognition and computer vision, see for example \cite{mizumachi2016superdirective, li2019vector, jiang2019deep}. The deep structure has a greater approximation power than a shallow neural network. There have been a few works regarding the approximation power of deep neural networks \cite{bianchini2014complexity, poggio2017and, shaham2018provable, fan2020universal}. The results suggest that using a deep neural network helps reduce the approximation error, which is useful in the cases where the approximation error dominates the total error. Therefore, it's necessary to consider a deep neural network model over a shallow model. However, finding a way to make the deep neural network well trained (statistically consistent) on small sample size is necessary. 

In this paper, we  discuss the stage-wise variable selection algorithm with neural networks. We  show that the existing stage-wise algorithm performs well at the beginning and selects the correct variables. However, at the later steps, the probability that it  selects a correct variable decreases. Thus we  propose an ensemble algorithm embedded with the stage-wise variable selection algorithm, named ensemble neural network selection (ENNS) algorithm. We  show that the new algorithm selects all correct variables with high probability and its false positive rate converges to zero.
Moreover, instead of a regular neural network trained on the selected variables, we propose a few methods to further reduce the variance of the final model. We provide some theoretical developments for statistical properties. Then we conduct an extensive numerical study to support our theoretical claims and validation of the proposed methodology. In the sequel, we propose an algorithm for the $l_1$ penalized neural network with soft-thresholding operators that addresses over-fitting issues.

The article is organized as follows: In section \ref{section:relatedwork}, we  discuss some major related works and the intrinsic dimensionality of a complex model. In section \ref{section:twostep}, we  present the ideas and algorithms behind the ENNS algorithm and the methods of increasing stability during the estimation step. In section \ref{section:methodology}, we  provide the theory for supporting the methodology.  In section \ref{section:simulation}, we  present our numeric study to support our claims.
Section \ref{section:conclusion}  concludes the article with some future directions.

\section{Related works}
\label{section:relatedwork}
In this section, we discuss some important  related works.

\subsection{The regularization approach  }
There is a considerable amount of  literature  that have discussed achieving sparsity via regularization \cite{tibshirani1996regression, fan2001variable, zou2006adaptive, yuan2006model, liu2007variable, fan2008sure, huang2010variable, marra2011practical}. Let $\btheta$ be the parameters of a model. A direct method is to add a zero norm of the parameters $\|\btheta\|_0$ to the loss function. However, optimizing a loss function with zero norms has been proved to be a non-deterministic polynomial-time hardness (NP-hard) problem, which requires exponential time to solve, \cite{natarajan1995sparse}. However, it has been proved that instead of directly penalizing the number of nonzero coefficients by $l_0$ norm, $l_1$ type penalty is able to shrink some coefficients to zero, and thus the features corresponding to these zero coefficients are not included in the model, \cite{tibshirani1996regression}. A great number of work has been done with the lasso estimator. Adaptive methods (\cite{zou2006adaptive}) assigns different weights to different coefficients using a data-driven way, which obtains oracle solutions. \cite{chatterjee2011bootstrapping} studied the bootstrap lasso estimator and its variations, and showed the strong consistency of the estimator. \cite{chatterjee2013rates} studies the consistency properties in the bootstrap adaptive lasso. \cite{das2019perturbation} proposed a modified perturbation bootstrap adaptive lasso estimator and showed its higher order consistency in the high-dimensional set up. The lasso penalty can be extended to adopt group-wise penalization, \cite{yuan2006model, huang2010variable}, commonly known as group lasso. The power of regularization is decided through the hyperparameter, which is also called the tuning parameter, by checking some criteria, such as,  BIC (\cite{schwarz1978estimating}), EBIC  (\cite{chen2008extended}), GIC (\cite{zhang2010regularization, fan2013tuning}) or by cross-validations.

Variations of the $l_1$ norm regularization are also widely used. It's known that $l_1$ norm regularization yields sparse solution (\cite{tibshirani1996regression}), while $l_2$ norm regularization controls the magnitude of coefficients, eases multicolinearity and overfitting. The $l_1+l_2$ norm regularization both yields sparse solution and encourages group effects (\cite{zou2005regularization}). The $l_{p,1}$ norm penalization (\cite{yuan2006model}, \cite{argyriou2007multi}), where $p=2$ matches the group lasso penalty, yields group sparsity. The $l_{p,1}, l_1$ norm penalty, known as the sparse group lasso penalty, yields both group sparsity and in-group sparsity solutions, (\cite{simon2013sparse}). Some variation of the norm regularization methods include the Dantzig selector \cite{candes2007dantzig}, which is a variation of the lasso. The SCAD \cite{fan2001variable} is defined from the derivative, instead of directly defined from the penalty term.

The regularization approach requires the assumption that only a few features are relevant in predicting the response, all other features either have exactly zero coefficients (strong sparsity assumption), for example, in the linear regression set up
$$\beta_j=0,\quad for\ j\in A^0\subset\{1,...,p\}\qquad and \qquad \#\{\beta_j,\ j\notin A^0\}=s<<p,$$
or the coefficients of irrelevant features are bounded from above by a negligible term (weak sparsity assumption)
$$\sum_{j\in A^0}|\beta_j|\leq\eta\qquad and\qquad \#\{\beta_j,\ j\notin A^0\}=s<<p$$
(for example, see \cite{zhang2008sparsity}). If the sparsity assumption is satisfied, the regularization approach with proper penalty yields a sparse solution with high probability (that the true subset of relevant features is selected ) under some common mild conditions. An advantage of this approach is that we know which features are selected, and thus the model has better interpretability.

\subsection{Deep neural network approximation}

It's well known,  the universal approximation theorem  \cite{cybenko1989approximation}, that a shallow neural network with $k$ hidden nodes, denoted as $SN_k(x)$ can be used to approximate any continuous function $f(x)$ defined in a bounded domain with arbitrarily small error
$$|SN_k(x)-f(x)|<\epsilon$$
for all $x$ in the bounded domain with a big enough $k$. This pioneering theorem encourages neural network to be used widely. Later as the development of deep neural networks, researchers found the limitations of shallow neural network such that the number of neurons needed to achieve a desire error increases as fast as exponentially \cite{chui1994neural, chui1996limitations}. Later, they  found that ``the two hidden layer model  may be significantly more promising than the single hidden layer model" \cite{pinkus1999approximation}. Sum neural networks, or equivalently, polynomial neural networks have been studied \cite{delalleau2011shallow, livni2013provably}, and universal approximation property has been established recently by \cite{fan2020universal} that a continuous function in $\mathcal{F}_d^n$ can be approximated with error $\epsilon$ by a quadratic network that has depth
$$O\left(\log(\log(\frac{1}{\epsilon}))+\log(\frac{1}{\epsilon})\right)$$
and number of weights
$$O\left(\log(\log(\frac{1}{\epsilon}))(\frac{1}{\epsilon})^{d/n}+\log(\frac{1}{\epsilon})(\frac{1}{\epsilon})^{d/n}\right)$$
where $d$ is the dimension of domain. The approximation theory for regular deep neural networks have also been established recently. \cite{poggio2017and} showed that a deep network need
$$O\left(\left(n-1\right)\left(\frac{\epsilon}{L}\right)^{-2}\right)$$
model complexity to approximate a $L$-Lipshitz continuous function of $n$ variables instead of
$$O\left(\left(\frac{\epsilon}{L}\right)^{-n}\right)$$
in a shallow neural network. \cite{shaham2018provable} and \cite{siegel2019approximation} provide much detail mathematical treatment for the deep neural network approximation power. 

\subsection{Variable selection and regularization in neural networks}

In terms of variable selection in neural networks, \cite{castellano2000variable} proposes an algorithm to prune hidden nodes in low-dimensional setup, while \cite{srivastava2014dropout} proposes a dropout technique to eliminate hidden nodes randomly. These methods set coefficients to zero and thus reduce the generalization variance, but do not help in the high-dimensional setup, where one needs to eliminate the unimportant input features.

In the high-dimensional setup, a neural network has even many more parameters and thus is harder to train compared to its low-dimensional setup. When we have a  relatively smaller sample size compared to the huge number of parameters, a neural network usually having high variance. A few researchers have studied this property \cite{feng2017sparse, liu2017deep, yang2020statistical, lemhadri2019neural}. By applying the group lasso regularization \cite{yuan2006model}, one can shrink the whole connections of a specific variable to exact zero, and thus performs variable selection. However, regularization methods in neural networks involve too many tuning parameters and therefore make the neural network sensitive to a small change in the tuning parameters in practice. \cite{liu2017deep} also presents a stage-wise variable selection algorithm with neural networks, called deep neural pursuit (DNP), which uses correlation to add new variables and enjoys faster speed. These methods are extensions of the high-dimensional linear models or additive models, which act as pioneers of this emerging topic.

\subsection{Algorithms}
The regularization methods usually involve $l_1$ norm penalty term, which is not easy to solve using regular gradient descent algorithms, see, for example, \cite{wright2015coordinate}. This issue is general for all models with $l_1$ penalty. The regularization path for generalized linear models can be easily obtained from coordinate descent algorithms \cite{wu2008coordinate, friedman2010regularization}.

Various algorithms are used to obtain a path selection. The least angle regression  provides a forward algorithm to add new features by looking at the correlation, \cite{efron2004least}. The LARS algorithm with simple modification can be used to obtain the lasso solution path. \cite{tibshirani2015general} provides a stage-wise algorithm, which provides a very close solution path to the lasso solution path. \cite{perkins2003grafting} studied a stage-wise algorithm to incorporate the $l_2$, $l_1$ and $l_0$ norm penalty with  gradients with respect to the input weights. The gradient has implicit connections with the correlation studied by \cite{efron2004least}.

For the other penalties, \cite{tewari2011greedy} shows that there is an equivalence between using the stage-wise algorithm and the group lasso penalty. \cite{liu2017deep} applies  deep artificial neural networks to perform feature selection.

\section{Proposed Methodology: Two-step variable selection and estimation }
\label{section:twostep}
In this section, we propose a two-step variable selection and estimation approach with deep neural networks. We  discuss the methodology in the following subsections. 

Just to recap, we replicate the data structure again. Consider a feature vector $\boldx\in\R^p$ and a response $y\in\R$ for the regression set up and $y\in\{0, 1\}$ in the classification set up. We have data $\{(\boldx_1,y_1),...,(\boldx_n,y_n)\}$ consisting of independent observations. Denote the design matrix $\boldX=(\boldx_1,...,\boldx_n)^T\in\R^{n\times p}$ and the response vector $\boldy=(y_1,...,y_n)^T$. As we mentioned before, we have more variables than observations, i.e.,  $p > n$. According to the previous discussion, variable selection is an important step in high-dimensional modeling. If one includes all variables in the model, there will be at least $p$ parameters to estimate, which can not be done consistently with the $n$ observations. If a more complicated model is needed, the number of parameters will be tremendous, which will cause severe over-fitting and high variance with a small training sample size.

Therefore, we hope a feature selection step at first can help pick the import variables, and another estimation step could build a more accurate model based on the selected variables. Moreover, we will use deep neural networks as the structure, since it will be able to capture the complicated relationships. We will consider a stage-wise algorithm in the variable selection step, which operates similarly to the DNP model, \cite{liu2017deep}. However, we will show that the stage-wise algorithm in DNP suffers from some disadvantage and propose an ensemble algorithm to relieve this situation. In the second step, we will discuss the methods of variance reduction and prevent over-fitting, since a deep neural network with only a few input variables can still have a huge number of parameters.

\subsection{The ensemble neural network selection (ENNS) algorithm}
Consider the feature selection approach in \cite{liu2017deep}. Let $\mathcal{D}:\R^p\rightarrow\R$ be a deep neural network  function that maps the original feature space to the output space. We don't specifically mark the number of hidden layers and hidden node sizes in notation, but simply assume the deep neural network has $m$ hidden layers with sizes $h_1,...,h_m$. Denote the weight matrices in each layer to be $\boldW_0,...,\boldW_m$, where $\boldW_0\in\R^{p\times h_1}$, $\boldW_i\in\R^{h_i\times h_{i+1}}$ for $i=1,...,m-1$ and $\boldW_m\in\R^{h_m\times 1}$. Denote $\boldt_i$ the intercept for the $i^{th}$ hidden layer and $b$ the intercept of the output layer. Let $\btheta = (\boldW_0,...,\boldW_m,\boldt_1,...,\boldt_m,b)$ be the parameters in the neural network model. For an input $\boldx\in\R^p$, denote the output
\begin{equation}
    \eta_{\btheta,\boldx} = \mathcal{D}_{\btheta}(\boldx)
\end{equation}
where in the regression set up, the output is from a linear layer and $\eta\in\R$, while in the classification, an extra sigmoid layer is added and $\eta\in(0,1)$. Moreover, we assume sparse feature, i.e., only a small fraction of the variables are significantly related to the response. Without loss of generality, we assume
$$\mathcal{S}_0=\{1,...,s\}$$
of the variables are truly nonzero variables.

Define the loss function for regression to be the squared error loss
\begin{equation}
    l(\btheta) = \mathbb{E}\left[(y-\eta)^2\right]
\end{equation}
In practice, we work with the empirical loss
\begin{equation}
    l(\btheta;\boldX,\boldy)=\frac{1}{n}\|\boldy-\bdeta\|_2^2
\end{equation}
where $\bdeta\in\R^n$ with $\eta_i=\eta_{\btheta,\boldx_i}$, $i=1,...,n$. Define the loss function for classification to be the negative log-likelihood, which is known as the cross-entropy loss
\begin{equation}
    l(\btheta)=\mathbb{E}\left[y\log\eta+(1-y)\log(1-\eta)\right]
\end{equation}
In practice, we work with the empirical loss
\begin{equation}
    l(\btheta;\boldX,\boldy) = \frac{1}{n}\sum_{i=1}^n\left[y_i\log\eta_i+(1-y_i)\log(1-\eta_i)\right]
\end{equation}

Let $\boldG_i$ be the gradient of the loss function with respect to $\boldW_i$ in the back propagation process for $i=0,...,m$, i.e.,
\begin{equation}
    \boldG_i=\frac{\partial}{\partial\boldW_i}l(\btheta;\boldX,\boldy),\quad i=0,...,m
\end{equation}
The DNP algorithm starts with the null model and adds one variable at a time. Let $\mathcal{S}$ be the selected set and $\mathcal{C}$ be the candidate set. At the beginning, we have
\begin{equation}
    \mathcal{S}=\{intercept\}\quad and\quad \mathcal{C}=\{1,...,p\}
\end{equation}
The model is trained on $\mathcal{S}$ only and the submatrix of $\boldW_0$ corresponding to the features in $\mathcal{C}$ is kept zero. After the training done, one chooses a $l_q$ norm (usually with $q=2$) and compute the gradient' norm for each $j\in\mathcal{C}$ of $\boldW_0$.
\begin{equation}
    \boldG_{0j} = \frac{\partial}{\partial \boldW_{0j}}l(\btheta;\boldX,\boldy),\quad j\in\mathcal{C}
\end{equation}
The next variable that enters the model, $j_+$ is
\begin{equation}
    j_+=\argmax_{j\in\mathcal{C}}\|\boldG_{0j}\|_q
\end{equation}
Then $\mathcal{S} = \mathcal{S}\cup \{j_+\}$ and $\mathcal{C} = \mathcal{C}/\{j_+\}$. 
To increase the stability, instead of computing $\boldG_{0j}$ directly, the DNP algorithm computes $\boldG_{0j}$ through the average over multiple dropouts. Let $B_1$ be the number of dropouts, the next variable is

\begin{equation}
    j_+ = \argmax_{j\in\mathcal{C}}\frac{1}{B_1}\sum_{b=1}^{B_1}\|\boldG_{b0j}\|_q
\end{equation}
where $\boldG_{b0j}$ denotes the gradient of the loss function with respect to the first layer's $j^{th}$ weight vector after the $b^{th}$ random dropout.

The algorithm works because $\|\boldG_{0j}\|_q$ describes how much the loss function will change when the next update on the corresponding variable's weight is performed, \cite{perkins2003grafting}. \cite{tewari2011greedy} also indicates that selecting variable by comparing $\|\boldG_{0j}\|_q$ has an equivalence to applying the group lasso penalization, see also \cite{liu2007variable}. 

The algorithm works well at the very beginning, which is described by proposition \ref{prop:begincompare} and proposition \ref{prop:beginprobability} in section \ref{section:methodology}. However, it suffers from a few disadvantages. First, as we include more correct variables in the model, the probability that we select another correct variable decreases. A simulation study in section \ref{section:simulation} also provides numeric support for this argument. Second, one needs to pre-specify how many variables need to be selected before stopping, denoted $s_0$. If this number is chosen to be more than the number of true variables, denoted $s$, there will be $s_0-s$ additional variables that should not have been included, i.e., the false positive rate could be high. Finally, the model does not use dropout or regularization during prediction, which has potential over-fitting risk. Here we propose the ensemble neural network selection (ENNS) algorithm to remove these issues, and we  discuss possible solutions in preventing over-fitting at the regression or classification step.

One could observe that when a fraction of $\mathcal{S}_0$ are already involved in the model, i.e., in $\mathcal{S}$, the model is trained such that these variables are used to explain the variations by all variables in $\mathcal{S}_0$. This weakens the effect of those truly nonzero variables in $\mathcal{C}$. These variables become less important than when there's no variable in $\mathcal{S}$. Moreover, there are less true nonzero and zero variables in $\mathcal{C}$ than at the very beginning, the probability that we select a correct variable in the next step is
\begin{equation}
    \prob(j_{next}\in\mathcal{S}_0) = \sum_{j\in\mathcal{S}_0\cap\mathcal{C}}\prob(j_{next}=j)
\end{equation}
which will be sequentially lower as $|\mathcal{S}_0\cap\mathcal{C}|$ decreases. Therefore, there will be a nonzero probability that at one stage the selected variable does not belong to $\mathcal{S}_0$. We consider an ensemble method to overcome this issue.
 
The idea behind this ensemble method is motivated by bagging, (\cite{breiman1996bagging, buhlmann2002analyzing}). Assume that we want to add $s_j$ variables in one step. Consider a bootstrap sample of size $n_b$ from the original sample. The DNP with random initialization is trained on this sample, which yields a selection set $\mathcal{S}_1=\{j_1,...,j_{s_j}\}$. Instead of just doing one pass, we propose that for $b_2$ in $1, ..., B_2$ and a bootstrap sample size $n_r$, we perform the feature selection on a random selection of $n_r$ observations. Denote the features being selected in all $B_2$ rounds as
$$\mathcal{J}_1 = \{j_{11}, ..., j_{1s_j}\}, ..., \mathcal{J}_{B_2} = \{j_{B_21}, ..., j_{B_2s_j}\}$$
We will only allow a variable to enter the model if it appears at least $[B_2p_s]$ times in the $B_2$ rounds, for a fixed proportion $p_s$. Mathematically,
$$\mathcal{J} = \{j\ in\ at\ least\ [B_2p_s]\ of\ \mathcal{J}_1, ..., \mathcal{J}_{B_2}\}$$
is the set of variables that will actually enter the model in this step.

The reason that this ensemble  improves the selection, lies on three points. First, the algorithm is an averaging of different bootstrapping results, thus the effect of some extreme observations could be averaged out. The final selection result represents the common part of the whole sample. Second, the neural network uses random initialization. In two different training, though the predictions seem similar, the estimated parameters are actually from different local minimums of the loss function. Therefore, these different training results represent different aspects of the model. Combining the two reasons, if we select a smaller $n_r$ compared to $n$, the selection results are closer to independent. However, $n_r$ can not be too small to avoid misleading the neural network. Finally, if a variable is selected by mistake in some round, this is possibly due to the specific bootstrap sample making the relationship between the variable and the response stronger, which is not general in all bootstrap samples. In practice, one will observe that though false selection happens, those false variables are different in different rounds. Therefore, this ensemble will actually make the probability of false selection tend to zero. This observation is supported by theorem \ref{thm:nofalsepositive} in section \ref{section:methodology}. Moreover, if two variables' interaction effect is important in the model, they are likely to be included in the model at the same step.

It's possible that the proposed method selects less or more variables than the number of variables we specified, $s_j$. If the sample is not enough to represent the true relationship between the variables and the response, it's very possible that the number of selected variables, denoted $\hat{s}_j$, is less than $s_j$. In this case, we exclude the variables that are already in $C$ from the neural network and perform another round of variable selection with $\mathcal{S}=\{1,...,p\}/\mathcal{C}$ and then $\mathcal{C}=\{intercept\}$. The number of variables to be selected in this round will be $s_j-\hat{s}_j$. On the other hand, $\hat{s}_j$ being more than $s_j$ happens when the selection proportion $p_s$ is specified too small. In this case, one would sort the variables by their appearing proportions and only select the first $s_j$ variables in the list.

In summary, we specify a number $s_0$ at the very beginning, which mean the final model will include $s_0$ variables. In the $j^{th}$ iteration, let $s_j$ be the number of variables to be selected. Right now there are $|\mathcal{S}_j|$ variables in the model, denoted $\mathcal{S}_{j}$. Let $\boldX_{-n}$ be the sub-matrix of $\boldX$ where the columns with indices in $\mathcal{S}_{j}$ removed. Train the ensemble on $\boldX_{-n}$ and obtain selection result $\hat{\mathcal{S}}_j$. Let $s_{j+1} = s_j - |\hat{\mathcal{S}}_j|$. The algorithm is repeated until the model has selected $s_0$ variables. An algorithm is given in Algorithm \ref{algorithm:selection}. Under mild assumptions, the algorithm  finally reaches selection consistency. This argument has been justified mathematically in theorem \ref{thm:selectionconsistency}, section \ref{section:methodology}. Moreover, a comparison of the variable selection performances of different modeling is also presented in section \ref{section:simulation}.

\begin{algorithm}

\SetAlgoLined
\hrulefill\\
Initialize number of selected variables $S=\emptyset$, $s=0$ and target $s_0$\;
\While{$|S|<s_0$}{
    \For{$b=1,...,B$}{
        Bootstrap sampling\;
        Random initialization with zero feature\;
        Run the DNP algorithm and obtain selection set $\mathcal{J}_b$\;
    }
    Obtain $\mathcal{J} = \bigcup_{b=1}^B \mathcal{J}_b$\;
    Compute $\mathcal{J}_T$ by filtering number of appearance\;
    \eIf{$\mathcal{J}_T <= s_0 - |S|$}{
        $S=S\cup \mathcal{J}_T$\;
        Remove the columns in $\mathcal{J}_T$ from training data\;
    }
    {
        $\mathcal{J}_T$ is the $m - s$ elements with highest number of appearance\;
        $S=S\cup \mathcal{J}_T$\;
    }
}
\hrulefill
\caption{Algorithm for feature selection ENNS}
\label{algorithm:selection}
\end{algorithm}

The computation complexity of the ENNS algorithm on a single machine is the number of bagged neural networks times the computation complexity of training a single neural network, which is equal to $O(Bhsnp)$. Here $B$ is the number of bagged neural networks, $h$ is the neural network structure complexity, $s$ is the number of variables to be selected, $n$ is the sample size and $p$ is the variable dimension. However, since bagging algorithm has independent elements, it's easy to parallelize the bagged neural networks by submitting different jobs. In this case, the computation complexity reduces to $O(hsnp)$, which is the same as that of DNP in \cite{liu2017deep}. As a comparison, \cite{liu2017deep} also mentioned the computation complexity of HSIC-Lasso in \cite{yamada2014high}, which grows cubicly with the sample size as $O(sn^3p)$.

Moreover, the algorithm is less sensitive to the choice of $s_0$ compared to the DNP (without ensemble) given that it is sufficiently large, . The ENNS algorithm will be more likely stopped when an appropriate number of variables enter. Specifying a too small $s_0$ will harm both the algorithms, since one forces the algorithm to stop before the correct number of variables enter. In practice, specifying a large $s_0$ to ENNS or tune $s_0$ with cross-validation works well. In detail, the cross-validation can be performed on the ENNS with a few pre-specified $s_0$'s and then the second stage estimation can be performed as described in the next subsection. 

\subsection{Estimation with regularization}
In this subsection ,we will discuss possible procedures to prevent over-fitting. After feature selection, the deep neural network can be trained on the selected features. However, the number of parameters in the neural network model is still huge. For example,  a  neural network with four hidden layers consist of $h_1,\cdots, h_4$ nodes respectively and $s$ selected variables  has $sh_1+h_1h_2+h_2h_3+h_3h_4+h_4$ parameters (without counting the intercepts). Thus, if one uses the  hidden layer sizes as $[50, 30, 15, 10]$ with the number of selected variables being five, that brings 2466 parameters. As a comparison, the linear model will have only six parameters, while a GAM with four knots and degree three will have 36 parameters. Therefore, fitting a deep network with a small sample size  still face  challenging issues. One has to be careful of  training  the neural network on the selected variables. A few possible methods are discussed below. The Xavier initialization  is used here to assure that the initial weights are in a proper range, \cite{glorot2010understanding}.

\subsubsection{Dropping out and bagging}

During the variable selection step, over-fitting is overcome by dropout layers, where we randomly set parameters to zero in the later layers. However,  dropout layers in prediction is risky, since we are not able to measure the prediction quality when performing a random dropout. One way out is,  use bagging again in this step. First, the connections in the estimated neural network, denoted $\mathcal{N}$ is randomly cut off, i.e., the weights are set to zero. By doing this we obtain a model $\mathcal{N}_r$, where $r$ stands for reduced connections. Then a prediction is made on model $\mathcal{N}_r$, denoted $\hat{y}_r$. This process is repeated for $K$ times. Denote the reduced neural networks to be $\mathcal{N}_{kr}$ and the associated predictions are $\hat{y}_{kr}$. In the regression set up, the final prediction is defined as
$$\hat{\boldy}=\frac{1}{K}\sum_{k=1}^K\hat{y}_{kr}$$

In the classification set up, the final prediction is defined as
$$\hat{y}_i = 
\begin{dcases}
1,\quad if\ \hat{p}_i > p_c\\
0,\quad if\ \hat{p}_i < p_c
\end{dcases}
,\quad for\ i=1,...,n
$$
where $p_c$ is some pre-specified threshold. In practice, a typical choice is $p_c=1/2$.
$$
\hat{p} = \frac{1}{K}\sum_{k=1}^K\hat{y}_{kr}
$$
and $\hat{p}_i$ is the $i^{th}$ element of $\hat{p}$. A simulation study is performed to show its efficacy in Section \ref{section:simulation}.

\subsubsection{Stage-wise training}

The stage-wise training idea comes from \cite{liu2017deep}, where the authors used it as a step-wise variable selection technique. However, here we adopt the idea to train the final model on the selected variables. The intuition behind this is that at each step, the information that is already trained remains in the training process. Therefore, adding a new variable adjusts the previous trained weights. Moreover, training with adaptive gradient algorithm (Adagrad, \cite{duchi2011adaptive}) allows adaptive learning rates for different parameters and thus ensures faster and more accurate convergence. In detail, assume that we have selected $m$ variables $\mathcal{J}$ from the ENNS algorithm. Let $\boldX_{\mathcal{J}}$ be the sub-matrix of $\boldX$ whose columns' indices are in $\mathcal{J}$. Then the DNP algorithm in \cite{liu2017deep} is trained on $\boldX_{\mathcal{J}}$ with $|\mathcal{J}|$ being the target number of variables. The performance is assessed through the simulation study reported  in Section \ref{section:simulation}.

\subsubsection{$l_1$ norm regularization}
As mentioned in section \ref{section:relatedwork} that $l_1$ regularization produces sparse neural network and controls over-fitting by shrinking parameters towards zero, and some parameters can be shrunk to exact zero.  Therefore, we choose to use $l_1$ norm regularization to control the parameter size and the number of nonzero parameters.

Let $\hat{S}$ be the set of indices of the variables that are selected from the first step. Let $\boldW=\boldW_1,...,\boldW_L$ be the hidden layer weights and $\boldt=t_1,...,t_L$ be the hidden layer intercepts (including the output layer). Let $f(x;\boldW,\boldt)$ be the neural network structure with such parameters that maps the original input to the output. In the regression problem, define
\begin{equation}
    \hat{\boldW},\hat{\boldt}=\argmin_{\boldW,\boldt}\frac{1}{n}\sum_{i=1}^n\left[y_i-f(\boldx_{\hat{S},i})\right]^2+\sum_{l=1}^L\lambda_{nl}|\boldW_l|,
\end{equation}
and in the classification problem, define
\begin{equation}
\label{equation:l1lossfunction}
\hat{\boldW},\hat{\boldt}=\argmin_{\boldW,\boldt}-\frac{1}{n}\sum_{i=1}^n\left[y_if(\boldx_{\hat{S},i})-\log(1+\exp(f(\boldx_{\hat{S},i})))\right]+\sum_{l=1}^L\lambda_{nl}|\boldW_l|,
\end{equation}
where $\boldx_{\hat{S},i}$ denotes the $i^{th}$ observation with only the selected variables. 

A direct training of the loss function (\ref{equation:l1lossfunction}) with the built-in $l_1$ loss penalty directly added to the cross-entropy loss does not work well in the current neural network libraries including tensorflow and pytorch. Therefore, a coordinate descent algorithm is needed to obtain sparsity in the neural network. Define the soft-thresholding operator $S(\cdot,\cdot):\R^d\times\R\rightarrow\R^d$ as
\begin{equation}
    \label{equation:softthresholding}
    (S(\boldx, c))_j=sign(x_j)(|x_j|-c)_+,\quad j=1,...,d.
\end{equation}

The algorithm consists an iterative process of updating the neural network weights without the $l_1$ penalty and then applying the soft-thresholding operator \ref{equation:softthresholding}. The number of epochs is pre-specified. However, the performance on the validation set can be monitored and an early-stopping criterion can be specified. The training will be stopped if the performance on the validation set does not improve for a pre-specified number of patience level. It worth noting that instead of selecting the tuning parameter, a sparsity level of each layer can be specified. Assuming there are $M$ hidden layers with sizes $h_1,h_2,...,h_M$. One may specify percentile $p_m$ for $m=1,...,M$. Denote $W_m$ the weight of layer $m$ and $W_{p_m}$ the $p_m^{th}$ percentile of $W_{p_m}$. Then for layer $m$, the soft-thresholding operator can be applied as $S(W_{m}, W_{p_m})$. For example, choosing a percentile of $50$ will make a  layer having $50\%$ sparsity level. An algorithm is given in Algorithm \ref{algorithm:l1estimation}. A simulation that compares the built-in $l_1$ penalty and the soft-thresholding operator is provided is Section \ref{section:simulation}.

\begin{algorithm}
\SetAlgoLined
\hrulefill\\
Initialize the weights with Xavier initialization\;
\While{Early stopping False OR epochs $<k$}{
    One step gradient descent for the neural network part\;
    \For{weights in layers}{
        Apply the soft-thresholding function with a pre-specified percentile\;
    }
    Check early stopping criterion\;
}
\hrulefill
\caption{Algorithm for $l_1$ norm estimation using coordinate descent}
\label{algorithm:l1estimation}
\end{algorithm}

\section{Theoretical Guarantee}
\label{section:methodology}
In this section, we develop theoretical supports for the proposed methodology and associated intuitions. A few assumptions are made in the derivations of  theoretical results. These assumptions along with two propositions can be found in the appendix \ref{section:assumptions}.

 Proposition \ref{prop:begincompare} and  \ref{prop:beginprobability} describe the behavior of neural network stage-wise algorithm at the very beginning. The probability that we select one predictor over another depends on the sum or the difference of their signal strengths. The greater the difference, the higher probability that we select the predictor with higher signal strength. The probability that we select a correct predictor at the very beginning is described by the error function and standard normal density functions. Though the form of the probability looks complicated,  the error function can be approximated by an exponential function with proper constants. Then the proposition shows that there is situation under which a non-zero variable entering into the model  with the stage-wise selection algorithm is not guaranteed with probability 1. Specifically, this happens when we have a low signal strength or a huge number of candidate variables.

So there is a concern that a wrong variable will mistakenly enter into the model due to a special training case of the neural network model, as shown in the previous proposition. With the bagging algorithm, we are able to eliminate the false positive selections with probability tending to 1. The intuition is that the false positive selection of a certain predictor happens due to a specific observation of the design matrix, which appears to be more correlated to the response or residual. However, with different sub-samplings, it's very unlikely that they yield the same wrong selection. This property is captured by the following theorem.
\begin{thm}
\label{thm:nofalsepositive}
Under the assumptions \ref{assumption:sparsitystrong} \ref{assumption:uncorrelatedness}, \ref{assumption:designmatrix} and \ref{assumption:sampingproportion}, and also assume that
$$s_0\leq C\cdot s=o(p)\quad and\quad \frac{p-s_0}{s}e^{-n\gamma_n}\rightarrow 0\ as\ n\rightarrow\infty.$$
Let $K_n$ be the estimated parameters' $l_1$ norm, which is assumed to satisfy
$$K_n^2\sqrt{\frac{\log(nK_n)}{n}}\rightarrow\infty\quad as\ n\rightarrow\infty,$$
then in each selection step of the ENNS algorithm, the probability of false positive converges to zero, i.e.
\begin{equation}
    \prob(j\in\hat{\mathcal{S}}\ and\ j\notin\mathcal{S})\rightarrow 0\quad as\ n\rightarrow\infty\ and\ B_2\rightarrow\infty
\end{equation}
\end{thm}
A proof is given in Appendix \ref{appendix}. In this variable selection algorithms, the most important property is ability of selecting the correct predictors consistently. Here we show that ENNS enjoys this property in the following theorem.
\begin{thm}
\label{thm:selectionconsistency}
Under assumptions \ref{assumption:sparsitystrong}, \ref{assumption:uncorrelatedness}, \ref{assumption:designmatrix} and \ref{assumption:sampingproportion}, let $K_n$ be the upper bound of the norm of the best parameters in the neural network model when $\mathcal{S}$ is included, and $K$ be the size of the first hidden layer, with the ensemble, if $\gamma_n$ satisfies
$$K\frac{\log(p-s)}{n}\log\left(1-\frac{1}{2}e^{-c\gamma_n^2}\right)\rightarrow 0\quad as\ n\rightarrow\infty$$
for some constant $c$, and
$$K_n^2\sqrt{\frac{\log(nK_n)}{n}}\rightarrow\infty\quad as\ n\rightarrow\infty$$
the probability that all nonzero variables are selected converges to 1, i.e., 
\begin{equation}
    \prob(\hat{\mathcal{S}}=\mathcal{S})\rightarrow 1\quad as\ n\rightarrow\infty\ and\ B_2\rightarrow\infty
\end{equation}
\end{thm}
A proof is given in Appendix \ref{appendix}. The theorem \ref{thm:selectionconsistency} shows that the true nonzero variables with strong enough signal, the algorithm is able to select all of them with probability tending to $1$. The conditions are not verifiable in practice, however, extensive numerical examples in section \ref{section:simulation} show that the ENNS algorithm reaches selection consistency easier than the other algorithms.

For the estimation step, there has been a few theoretical works about the asymptotic properties, e.g.,   \cite{feng2017sparse, yang2020statistical}, where the results under sparse group lasso penalty are derived. The $l_1$ norm penalty in the estimation step is actually a special case of the sparse group lasso \cite{simon2013sparse} with the lasso weight $\alpha=1$ and the group lasso weight $1-\alpha=0$. Therefore, the results of these papers hold as long as we have $\hat{\mathcal{S}}=\mathcal{S}_0$, which has probability tending to $1$ by theorem \ref{thm:selectionconsistency}. Here we  adapt  \cite{gyorfi2006distribution} and  provide the following result. 

\begin{thm}
\label{thm:consistency}
Assume the assumptions \ref{assumption:sparsitystrong}, \ref{assumption:uncorrelatedness}, \ref{assumption:designmatrix} and \ref{assumption:sampingproportion}, consider the variables selected by the ENNS algorithm and the estimation with  $l_1$ regularization method. Denote the $l_1$ regularization tuning parameter by $\lambda_n$ and the corresponding Lagrangian parameter $K_n$. Denote the hidden layer size with $k_n$. In the regression set up, assume $\E(Y^2)<\infty$, if $K_n\rightarrow\infty$, $k_n\rightarrow\infty$ and $k_ns^2K_n^4\log(k_nsK_n^2)/n\rightarrow 0$, we have
$$\lim_{n\rightarrow\infty,B_2\rightarrow\infty}\prob\left(\E\int|f_n(\boldx)-f(\boldx)|^2\mu(dx)\rightarrow 0\right)=1$$
where $f_n$ is the estimated neural network and $f$ is the true function. In the classification set up, assuming that the probability of response being $1$ is bounded away from $0$ and $1$ by $\tilde{\epsilon}$, denote with $Q$, the maximum number of equivalent neural network classes, choosing tuning parameter $\lambda_n\geq c\sqrt{k_n\log n/n}(\sqrt{\log Q}+\sqrt{k_n\log s}\log(nk_n)$, if $\log(n)/(n\tilde{\epsilon}^2)\rightarrow 0$, $s^2k_n\lambda_n^2/(n\tilde{\epsilon}^2)\rightarrow 0$ and $n^{-1}k_n^{9/2}s^{5/2}\sqrt{\log(s_n)}\rightarrow 0$ as $n\rightarrow\infty$, we have
$$\lim_{n\rightarrow\infty,B_2\rightarrow\infty}\prob\left(R(f_n)-R(f^*)\rightarrow 0\right)=1$$
where $R(f_n)$ is the risk of neural network classifier and $R(f^*)$ is the risk of Bayes classifier.
\end{thm}
Theorem \ref{thm:consistency} states that under the previously discussed conditions, the regression reaches weak estimation consistency of the non-parametric function defined in \cite{gyorfi2006distribution}. For the classification, the neural network classifier's risk tends to the optimal risk, Bayes risk, see for example \cite{devroye2013probabilistic}. The theorem is a direct result from the existing results of the low dimension neural network regression model and classifiers, \cite{feng2017sparse, yang2020statistical}. Conditioning on the fact that we can select all correct variables with probability tending to $1$, applying the full probability formula, the consistency of the two-step approach can be derived with the low dimensional consistency plus the probability of non-selection-consistency.

The consistency error comes from two aspects: the variable selection error and the estimation error. The intuition behind this is that with a wrong selection result, the estimation error may be big, however, this happens with probability tending to zero. With a correct selection result, the estimation error behaves the same as in the low dimensional case, which converges to zero.

\section{Simulation study}
In this section, we use a few examples as numerical supports to our arguments in the previous sections. The code for DNP is composed according to the algorithm outline in \cite{liu2017deep}, and the code of ENNS is composed according to the algorithm in this paper. Both codes can be found at 
\url{https://github.com/KaixuYang/ENNS}
.

\label{section:simulation}
\subsection{Stage-wise correct selection probability decreasing study}
In this subsection, we use an example to demonstrate that the chance of selecting a correct variable in a stage-wise neural network decreases as we have more correct variables entering into the model. Consider a design matrix $\boldX$ that is drawn from a uniform distribution $(-1,1)$. The sample size is set to $n=1000$ and the number of predictors is set to $p=10000$. Let $s=5$ of the predictors are related with the response. We consider three different true structures of the relationship between the predictors and the response: linear, additive non-linear and neural network. For the response, we consider two different cases: regression and classification. In the linear case, the coefficients are drawn from a standard normal distribution. In the additive non-linear case, the functions are set to
\begin{equation}
    \eta = \sin(x_1)+x_2+\exp(x_3)+x_4^2+\log(x_5+2) - 2
\end{equation}
where $y=\eta+\epsilon$ in the linear case and $prob=\sigma(\eta)$ in the classification case. In the neural network case, we set hidden layers as $[50, 30, 15, 10]$ and weights from standard normal distribution.

For each of the cases, we test with starting from 0 to 4 correct predictors. In order to eliminate the effect of different signal strength from different predictors, we randomly sample $j$ indices from $1,...,5$ as the selected variables, for $j=0,...,4$, and include those $j$ indices predictors as the initially selected variables. We repeat the process 1000 times and report the proportion that the next variable that enters the model is a correct predictor. The result is summarized in table $\ref{table:probabilitydecrease}$.

\begin{table}
\caption{The proportion of correct variable selection after 0-4 correct variables in the model, for different cases over 1000 repetitions. The results show the mean.}
    \centering
    \fbox{%
    \begin{tabular}{ccccccc}
    \hline
    Response&structure&0 variable&1 variable&2 variables&3 variables&4 variables\\
    \hline
    \multirow{3}{*}{Regression}&Linear&0.995(0.002)&0.952(0.006)&0.863(0.010)&0.774(0.013)&0.430(0.016)  \\
         &Additive&0.993(0.003)&0.847(0.011)&0.905(0.009)&0.794(0.012)&0.531(0.016) \\
         &NN&0.998(0.001)&0.971(0.005)&0.932(0.007)&0.788(0.013)&0.574(0.016) \\
         \hline
         \multirow{3}{*}{Classification}&Linear&0.989(0.003)&0.918(0.009)&0.873(0.009)&0.813(0.011)&0.552(0.016)\\
         &Additive&0.992(0.003)&0.957(0.006)&0.911(0.009)&0.706(0.014)&0.633(0.015)\\
         &NN&0.994(0.002)&0.968(0.006)&0.947(0.004)&0.895(0.009)&0.762(0.013)\\
         \hline
    \end{tabular}}
    \label{table:probabilitydecrease}
\end{table}

From the table  $\ref{table:probabilitydecrease}$, we see that the probability of selecting a correct predictor decreases as  correct predictors enter into the model stage-wise, in all cases. The only exception is in the regression set up with additive non-linear structure from 1 variable to 2 variables, which could be due to random error.

\subsection{False positive rate study}
In this subsection, we  demonstrate empirically  that the false positive rate of the proposed ENNS (the probability of selecting a wrong variable) is superior than the pure stage-wise algorithm. Note that if one fixes the number of variables to be $s$, stage wise algorithm always select $s$ variables, while ENNS will stop if there isn't any new variable that satisfy the condition to be added. Therefore, it's possible that ENNS selects less number of variables and avoid wrong selection. We use the same setup as \cite{liu2017deep} to generate responses. Two different types of responses including regression and classification are considered here. The input variable $\boldX$ is drawn from $U(-1, 1)$, where the feature dimension $p$ is fixed at $10,000$. The corresponding labels are obtained by passing $\boldX$ into the feed forward neural network with hidden layer sizes $\{50, 30, 15, 10\}$ and ReLU activation functions. Input weights connecting the first $s$ inputs are randomly sampled from $N(1,1)$ for regression and $N(0,1)$ for classification. The remaining input weights are kept at zero. For each $s=2, 5, 10$, we generate 1000 training samples. In table \ref{table:falsepositive}, we report the false positive rate for the ENNS algorithm and the neural network stage-wise algorithm.

\begin{table}
\caption{Selection false positive rate average of the ENNS and DNP under different number of true variables in 101 repetitions. Standard deviations are given in parenthesis.}
    \centering
    \fbox{
    \begin{tabular}{ccccc}
    \hline
         Response&Method&s=2&s=5&s=10\\
         \hline
         \multirow{2}{*}{Regression}&ENNS&10.4\%(21.5\%)&11.5\%(22.1\%)&12.8\%(23.6\%)\\
         &DNP&22.5\%(29.5\%)&30.2\%(28.7\%)&41.4\%(33.2\%)\\
         \hline
         \multirow{2}{*}{Classification}&ENNS&4.7\%(17.9\%)&7.4\%(18.6\%)&9.8\%(20.3\%)\\
         &DNP&16.5\%(24.4\%)&24.8\%(29.7\%)&40.5\%(32.8\%)\\
         \hline
    \end{tabular}}
    
    \label{table:falsepositive}
\end{table}

Note that the ENNS's false positive rate is significantly less than that of DNP under significance level $\alpha=0.05$ (using the two-proportion Z test). The results provides strong evidence that the ENNS is effective in reducing the probability of selecting an incorrect variable.

\subsection{Variable selection simulation study}
In this subsection, we study the variable selection capability of our ensemble neural network selection (ENNS) algorithm in complex modelling. We retain similar setup as in the last subsection to generate responses. A slight difference is that the input weights connecting the first $s$ inputs are randomly sampled from $N(0,2)$ for regression and $N(0,1)$ for classification. The difference in the normal distribution parameters is to ensure similar signal-to-noise ratio (SNR) and ensures the signal strengths are in proper ranges for demonstration. The LASSO \cite{tibshirani1996regression} is implemented by the scikit learn library, and the HSIC lasso \cite{yamada2014high} is implemented using the HSICLasso library. In all four algorithms, the number of selected variables are strictly restricted to the true number of variables. In  ENNS, we run a bagging of 10 rounds with selection proportion 0.3. The Table \ref{table:selection} reports the average number of correct variables that are selected on 101 repetitions of the data. 

\begin{table}
 \caption{Variable selection capacity of ENNS and other methods with low signal strength in the regression (top) and classification (bottom) set up. The numbers reported are the average number of selected variables which are truly nonzero. The standard errors are given in parenthesis.}
    \centering
    \fbox{
    \begin{tabular}{ccccc}
    \hline
         Response&Method&s=2&s=5&s=10\\
         \hline
         \multirow{4}{*}{Regression}&ENNS&1.73(0.52)&4.21(0.56)&9.25(1.11)\\
         &DNP&1.61(0.50)&3.92(0.56)&8.77(1.13)\\
         &LASSO&1.65(0.57)&3.87(0.62)&9.62(1.38)\\
         &HSIC-LASSO&1.67(0.47)&3.80(0.66)&3.61(1.17)\\
         \hline
         \multirow{4}{*}{Classification}&ENNS&1.81(0.49)&4.24(0.87)&8.04(1.25)\\
         &DNP&1.67(0.76)&3.76(1.06)&5.95(1.29)\\
         &LASSO&1.67(0.56)&3.76(0.75)&5.76(1.38)\\
         &HSIC-LASSO&1.67(0.47)&2.80(0.91)&3.61(1.17)\\
         \hline
    \end{tabular}}
    \label{table:selection}
\end{table}

We observe that the ENNS outperforms the other variable selection algorithms in all three cases, and the difference is significant when $s=10$ under a t-test. The ENNS performs better when there are more nonzero variables. None of the algorithms were able to reconstruct the original feature indices due to a few reasons: the sample size is relatively small compared to the number of variables; the data generation through neural network structures is complicated; the signal strength is relatively low.

To fully study the variable selection power of the ENNS algorithm, we implemented another simulation in classification using higher signal strengths while keeping all other conditions the same. In the simulation study, we increase the mean weights of the nonzero variables to 3.5 and 10. With same implementations, we summarize the results in Table \ref{table:selection2}.

\begin{table}
    \caption{Variable selection capacity of ENNS and other methods with normal and high signal strength. The numbers reported are the average number of selected variables which are truly nonzero. The standard errors are given in parenthesis.}
    \centering
    \fbox{
    \begin{tabular}{cccccccc}
    \hline
    &\multicolumn{3}{c}{Normal Signal Strength}&\multicolumn{3}{c}{High Signal Strength}\\
    \hline
         Method&s=2&s=5&s=10&s=2&s=5&s=10\\
         \hline
         ENNS&2.00(0.00)&4.71(0.55)&8.38(2.06)&2.00(0.00)&5.00(0.00)&9.90(0.29)\\
         DNP&1.86(0.35)&4.38(0.84)&7.43(2.36)&2.00(0.00)&5.00(0.00)&9.47(1.10)\\
         LASSO&1.81(0.39)&4.19(1.01)&7.47(2.40)&2.00(0.00)&4.90(0.29)&9.23(1.74)\\
         HSIC-LASSO&1.71(0.45)&3.71(1.12)&4.95(2.13)&2.00(0.00)&4.62(0.84)&7.76(2.76)\\
         \hline
    \end{tabular}}

    \label{table:selection2}
\end{table}

The ENNS reaches selection consistency when $s=2$, while the other competitive algorithms still do not have selection consistency. However, all algorithms have obvious improvements in all cases under strong signals. We have to admit that selecting the correct subset of variables in all 101 repetitions is extremely challenging, since the data have great variability in different repetitions. Moreover, when $s$ becomes larger, the importance of a few variables are less likely to be observed from the data.

We like to reiterate that the bagging algorithm can be paralyzed since different runs are independent to each other. Therefore, the computational efficiency of this variable selection algorithm is almost the same as the computation efficiency of a single run.

\section{Real data examples}
\label{section:realdata}
In this section, we evaluate the performance of the two-step model on real-world data sets.

\subsection{Regression: riboflavin production data}
In this example, we consider the riboflavin production with bacillus subtilis data, which is publicly available in the `hdi' package in R. The data set contains a continuous response, which is the logarithm of the riboflavin production rate, and $p=4088$ predictors which are the logarithm of the expression level of 4088 genes. There are $n=71$ observations available. All predictors are continuous with positive values.

We perform 50 repetitions of the following actions. The data is split into training (56) and testing (15) observations. The training data is further split into training (49) and validation (7). The training data is normalized with mean zero and standard deviation one. We train the ENNS algorithm to select variables and perform the $l_1$ neural network to make a prediction. Along with our proposed algorithms, we compare the performance with the lasso penalized linear regression, which is implemented by the scikit-learn library in python; the group lasso penalized generalized additive model in \cite{yang2018ultra}, where the code can be found at 
\url{https://github.com/KaixuYang/PenalizedGAM}
; and the sparse group lasso penalized neural network in \cite{feng2017sparse}. Figure (\ref{figure:riboflavin}) shows the average testing mean squared error (MSE) along with the number of selected features for different models. Our proposed algorithm converges fast and produces competitive performance. Table \ref{table:riboflavinresult} shows the average prediction accuracy with standard error in parentheses and the median number of variables selected. Our proposed method has mean competitive performance but lowers standard errors.

The final model of small sample utilizes only two hidden layers, with over 90\% sparsity to prevent over-fitting, which is necessary for this small training sample size, 49. Training with large batch size, small learning rate, a huge number of epochs and early stopping help the model learn better and prevent over-fitting. We admit that tuning the network structure and learning parameters are hard, but we obtain better and stable results once we have the right numbers.

\begin{table}
\caption{Test MSE with standard error in parentheses and median of number of features for different models in the riboflavin gene data example.}
\centering
\fbox{
\begin{tabular}{ccc}
\hline
Model&Test MSE&Number of features\\
\hline
ENNS$+l_1$ neural network&0.268(0.115)&42\\
Regularized neural network&0.273(0.135)&44\\
Linear model with LASSO&0.286(0.124)&38\\
Generalized additive model with group lasso&0.282(0.136)&46\\
\hline
\end{tabular}}
\label{table:riboflavinresult}
\end{table}

\subsection{Classification: prostate cancer data}
In this example, we consider a prostate cancer gene expression data, which is publicly available at \url{http://featureselection.asu.edu/datasets.php}. The data set contains a binary response with 102 observations  on 5966 predictor variables. Clearly, the data set is really a high dimensional data set. The responses have values 1 (50 sample points) and 2 (52 sample points), where 1 indicates normal and 2 indicates tumor. All predictors are continuous predictors, with positive values.

\begin{figure}
    \centering
    \includegraphics[width=0.45\textwidth]{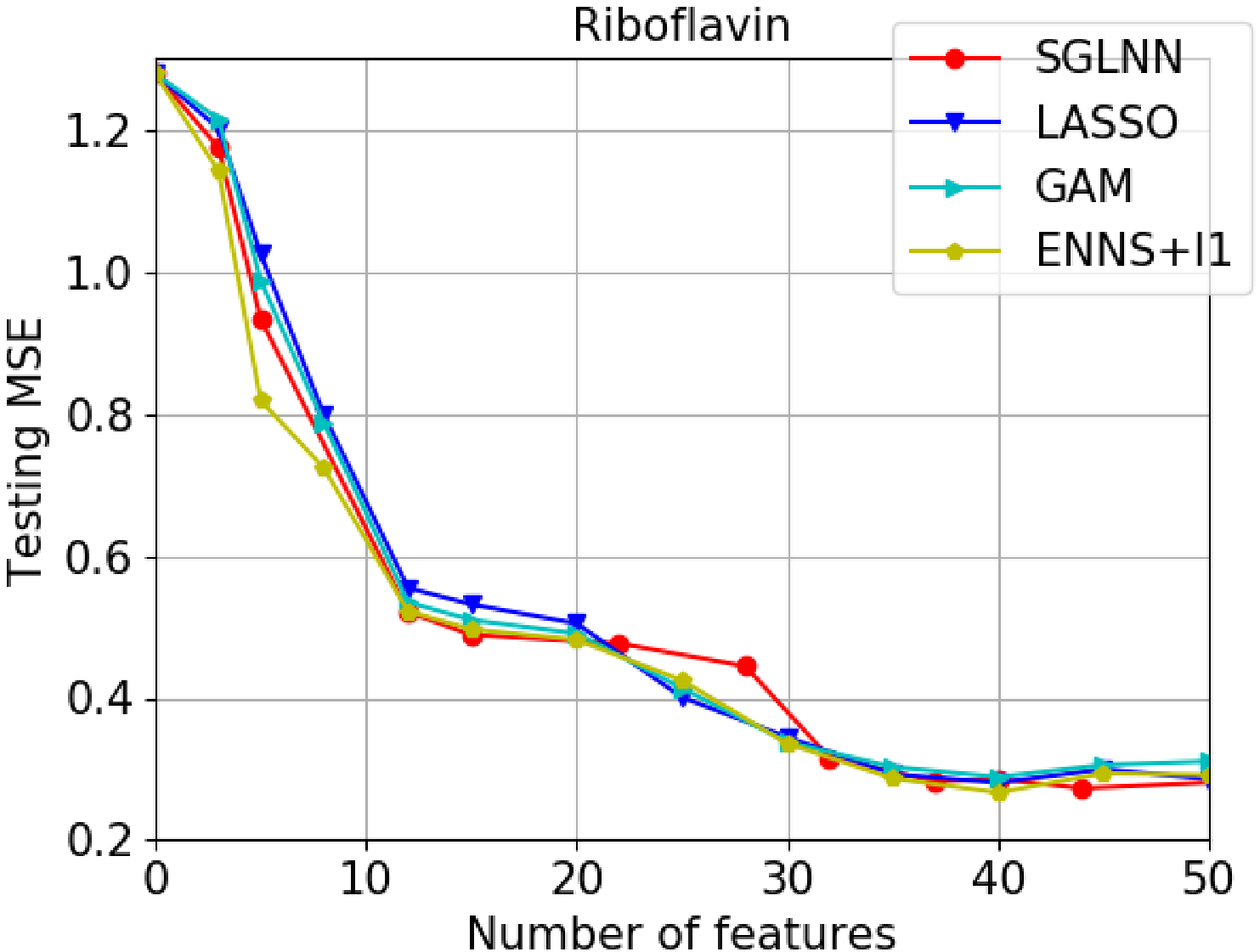}
    \includegraphics[width=0.45\textwidth]{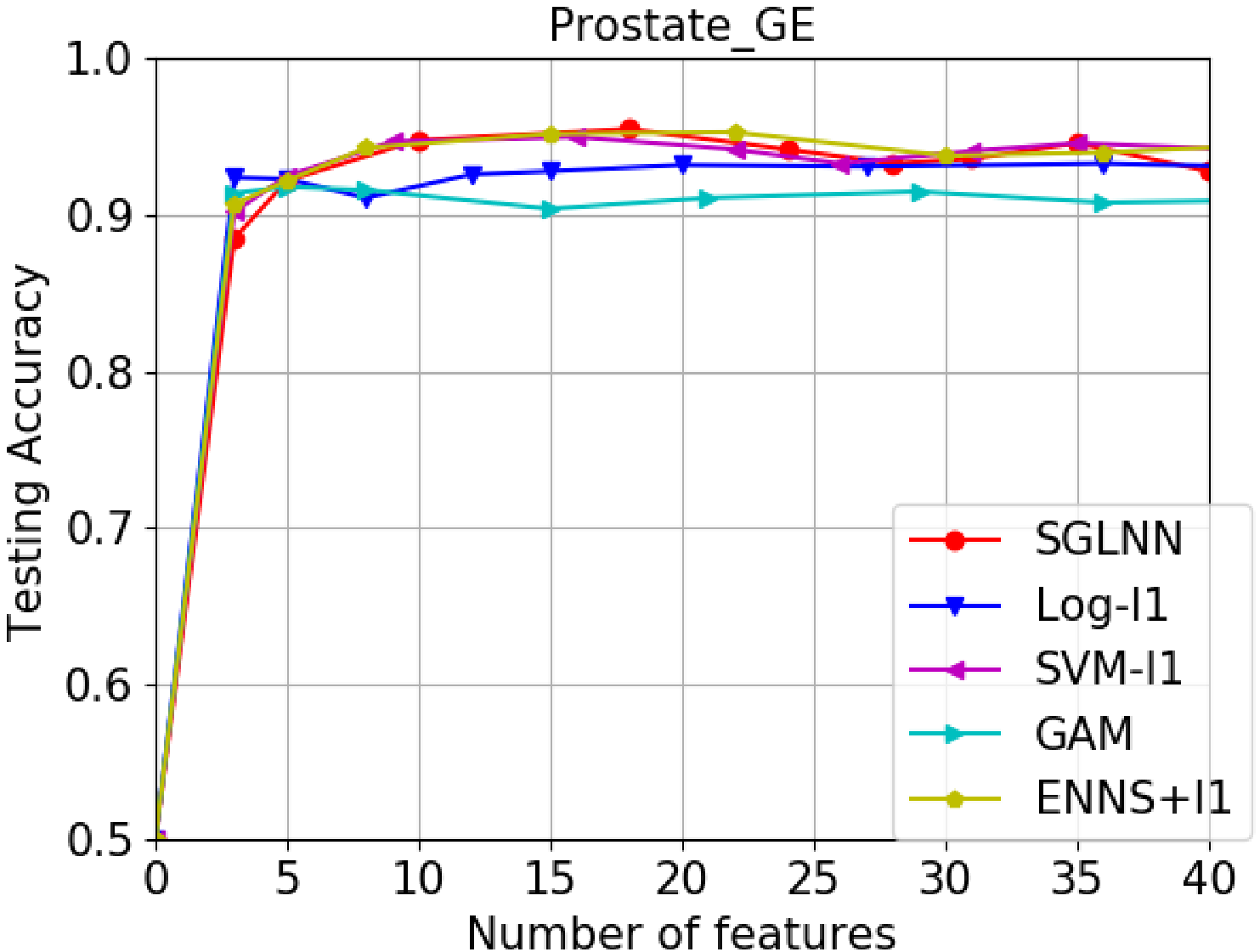}
        \caption{Testing mean squared error (MSE) for different models on the riboflavin data (left) and testing accuracy for different models on the prostate cancer data (right).}
    \label{figure:riboflavin}
\end{figure}

We perform 50 repetitions of the following actions. The data is split into training (81) and testing (21) observations. The training data is further split into training (70) and validation data (11). In each split, the number of class 0 observations and number of class 1 observations are kept roughly the same. We train the ENNS algorithm to select variables and perform the $l_1$ neural network to make predictions. Along with our proposed algorithms, we compare the performance with the $l_1$ norm penalized logistic regression; the $l_1$ support vector machine (SVM), both of which are implemented with the scikit-learn library in python; the group lasso penalized generalized additive model in \cite{yang2018ultra}, where the code can be found at 
\url{https://github.com/KaixuYang/PenalizedGAM}
; and the sparse group lasso penalized neural network of \cite{feng2017sparse}. Figure (\ref{figure:riboflavin}) shows the average testing accuracy over the 20 repetitions along with the number of selected features for different models. Our proposed algorithm converges fast and performs competitively. Table \ref{table:prostateresult} shows the average prediction accuracy with standard error in parentheses, and the median number of selected variables. Our proposed method has competitive mean performance but lower standard error. One needs to notice that the mean performance is hard to improve further, since the results are already good and reach the bottleneck of the current explanatory power. The reason that GAM performs worse than the other models is that the range of predictor variables are relatively small and skewed, thus the basis expansion on GAM does not work well.

\begin{table}
\caption{Test accuracy with standard error in parentheses and median of number of features for different classifiers in the Prostate gene data example.}
\centering
\fbox{
\begin{tabular}{ccc}
\hline
Classifier&Test accuracy&Number of features\\
\hline
ENNS$+l_1$ neural network&0.956(0.053)&15\\
Regularized neural network&0.955(0.066)&18\\
Logistic Regression with Lasso&0.933(0.058)&36\\
l1 penalized Linear SVM&0.950(0.052)&16\\
Generalized additive model with group lasso&0.918(0.061)&5\\
\hline
\end{tabular}}

\label{table:prostateresult}
\end{table}

\section{Conclusion}
\label{section:conclusion}

In this paper, we discussed the existing methods to deal with high-dimensional data and how to apply the stage-wise algorithm with neural networks. We discussed the shortcomings of the current stage-wise neural network variable selection algorithms and proposed a new algorithm,  ENNS to overcome these issues. We also  compared different methods to further reduce the over-fitting problem after the variable selection is done. Various theoretical derivations were provided to support the methodology, intuition, new algorithms, and extensive simulation studies were presented as empirical evidence that the methodology works in practice.
f
Though there's a few algorithmic and methodology studies with neural network variable selection, the theory for a neural network still deserves much more investigations. We hope this paper could work as a pioneer and attracts more  attention to the theory and methodology for high dimensional neural network models.

\section{Acknowledgement}
This research is partially supported by NSF-DMS 1945824.
\bibliography{ref}
\bibliographystyle{abbrv}

\newpage
\begin{center}
    {\LARGE Supplementary material: ENNS: Variable Selection, Regression, Classification and  Deep Neural Network for High-Dimensional Data}
\end{center}

\begin{appendix}
\section{Assumptions and propositions}
\label{section:assumptions}
\subsection{Assumptions}
The first  well-known assumption in high-dimensional modeling is sparsity. Here we provide two versions of sparsity: a stronger version and a weaker version. These are stated in assumption \ref{assumption:sparsityweak} and assumption \ref{assumption:sparsitystrong} respectively.

\begin{assumption}[Sparsity (weak)]
\label{assumption:sparsityweak}
The features are sparse, i.e., only $s<n=o(p)$ of the $p$ variables are strongly related with the response. Specifically, if $y$ depends on $x$ through a linear relationship with coefficients $\bbeta=(\beta_1,...,\beta_p)^T$, we have
\begin{equation}
    \min_{j=1,...,s}|\beta_j|\geq\gamma_n\qquad and\qquad \sum_{j=s+1}^p|\beta|_j=\tau_n=o(\gamma_n)
\end{equation}
where $\gamma_n$ is a sequence that may go to zero as $n$ goes to infinity.
\end{assumption}

\begin{assumption}[Sparsity (strong)]
\label{assumption:sparsitystrong}
The features are sparse, i.e., only $s<n=o(p)$ of the $p$ variables are related with the response. Specifically, if $y$ depends on $x$ through a linear relationship with coefficients $\bbeta=(\beta_1,...,\beta_p)^T$, we have
\begin{equation}
    \min_{j=1,...,s}|\beta_j|\geq\gamma_n\qquad and\qquad \sum_{j=s+1}^p|\beta|_j=0
\end{equation}
where $\gamma_n$ is a sequence that may go to zero as $n$ goes to infinity.
\end{assumption}

We know that in most cases the predictors are dependent, or at least weak correlation exists. However, even a weak correlation brings great complexity in theory. Therefore, we assume independent predictors in this section when deriving our theoretical results. However,  extensive simulation studies are performed to show the efficacy of our methodology  when the predictors are correlated. Further, though the theoretical results are proved under this independent assumption, they can be extended to the assumption that the correlation is up to $o(1)$.

\begin{assumption}[Independence]
\label{assumption:uncorrelatedness}
The predictors in the design matrix satisfy
\begin{equation}
    cor(x_j,x_k)=0,\ 1\leq j<k\leq p
\end{equation}
\end{assumption}

The behavior of the design matrix should also be controlled. Here we consider a random design and assume the following assumption
\begin{assumption}[Design matrix]
\label{assumption:designmatrix}
The covariate vector $\boldx$ has a continuous density and there exist constants $C_1$ and $C_2$ such that the density function $g_j$ of $\boldx_j$ satisfies $0 < C_1 \leq g_j (x) \leq C_2 < \infty$ on $[a,b]$ for every $1 \leq j \leq p$.
\end{assumption}

As a typical assumption for bagging, we require the bagging sample proportion to be not too small.
\begin{assumption}[Sample proportion]
\label{assumption:sampingproportion}
In each bagging round, every sample has $q_n$ probability to be included, where $q_n$ satisfies
$$nq_n\rightarrow\infty\quad as\ n\rightarrow\infty$$
\end{assumption}
Note that if we choose the bootstrap sample size to be the same as the sample size $n$, by law of large numbers, we have approximately $q_n=1-1/e$, where $e$ is the natural number.

\subsection{Propositions}
The following two propositions considers a scenario that the true underlying relationship between the predictors and the response is linear, which demonstrates how the probability of choosing one variable over another in the first step is decided. The first proposition gives the probability that we select one variable over another, and the second proposition gives the probability that we will select a correct variable in the first step. 

\begin{prop}
\label{prop:begincompare}
Consider the case where $y$ depends on $x$ through a linear structure with coefficients $\beta_1,...,\beta_p$ which satisfy assumption \ref{assumption:sparsityweak}. Also under assumption \ref{assumption:uncorrelatedness} and \ref{assumption:designmatrix}, if the sub-matrix of $\boldx$ consisting of the columns corresponding to the nonzero coefficients is column-wise orthogonal, let $c_j$ be the criterion to select predictor $j$. Recall that we will select predictor $j$ if $j=\argmax_jc_j$, where $c_j$ is the $L_2$ norm of the gradient with respect to the $j^{th}$ input. Then we have
\begin{align}
    \label{equation:begincompare}
    \prob(c_j<c_k)=&2L\left(\frac{|\beta_j|-|\beta_k|}{\sqrt{2}\sigma}, -\frac{|\beta_k|}{\sigma}, \frac{1}{\sqrt{2}}\right) + 2L\left(\frac{|\beta_j|+|\beta_k|}{\sqrt{2}\sigma}, \frac{|\beta_k|}{\sigma}, \frac{1}{\sqrt{2}}\right) + \nonumber\\
    &\Phi\left(\frac{|\beta_j|-|\beta_k|}{\sqrt{2}\sigma}\right) + \Phi\left(\frac{|\beta_j|+|\beta_k|}{\sqrt{2}\sigma}\right) - 2
\end{align}
where
\begin{equation}
    L(a, b, \rho)=\prob(X_1>a, X_2>b)
\end{equation}
is the bivariate orthant probability with correlation $\rho$ and $\Phi(\cdot)$ is the standard normal distribution CDF.
\end{prop}

\begin{prop}
\label{prop:beginprobability}
Under the assumptions \ref{assumption:sparsityweak}, \ref{assumption:uncorrelatedness} and \ref{assumption:designmatrix}, the probability that we select a nonzero predictor at the very beginning using the stage-wise neural network selection is
\begin{equation}
    \label{equation:beginprobability}
    \prob(\text{A nonzero predictor enters the model first})=\sum_{k=1}^s\int_0^{\infty}f_k(x)\prod_{j\neq k}^pF_j(x)dx
\end{equation}
where
\begin{align}
    &F_k(x)=\frac{1}{2}\left[erf\left(\frac{x+|\beta_k|}{\sqrt{2\sigma^2}}\right)+erf\left(\frac{x-|\beta_k|}{\sqrt{2\sigma^2}}\right)\right]\quad and\nonumber\\
    &f_k(x)=\frac{\partial}{\partial x}F_k(x)=\sqrt{\frac{2}{\pi\sigma^2}}e^{-\frac{x^2+\beta_k^2}{2\sigma^2}}\cosh{\frac{\beta_k x}{\sigma^2}}
\end{align}
and $erf(\cdot)$ is the error function. Moreover, if $\beta_{max}=\max_{j=1,...,s}$ is bounded, as $s\rightarrow\infty$, the probability is bounded from above
$$\prob\leq 1-\delta$$
where $\delta$ is a nontrivial quantity.
\end{prop}

The proofs of these two propositions are given in Appendix \ref{appendix}.

\section{Extra Numeric Results}
In this section, we present extra simulation results.

\subsection{Estimation simulation study}

In this subsection, we compare the estimation methods described in section \ref{section:twostep}. To fully understand the estimation differences among the competitive methods, we assume correct variable selection and perform the estimation on the correct subset of variables. The data are generated according to the same scheme as in the last subsection. We  compare the performance of these different estimation methods for $s=2, 5, 10$ assuming that we know the correct subset of variables. The simulation is run on 101 repetitions of data generation using different seeds. In the Table \ref{table:estimationcompare}, we report the RMSE, the MAE and the MAPE for regression, and the accuracy, the auc score and the f1 score for classification. 

On average, we see $l_1$ norm regularization provides best performance, except for the MAPE when $s=10$ in regression. Moreover, we observe that both built-in $l_1$ and soft-thresholding gives smaller standard errors, which coincides with the shrinkage estimator's properties. However, soft-thresholding provides better performance on average than built-in. The reason is that sparsity is not well supported with most libraries, thus a manual operation is needed to obtain sparsity.

\subsection{Variable selection and estimation}

In this subsection, we study the prediction capability of the  ENNS algorithm with $l_1$ neural network, and compare it with the DNP model, the logistic regression and the HSIC-lasso with SVM. We use the same neural network structure to generate data as in the previous section. Under 101 repetitions, we report the average RMSE (root mean squared error), MAE (mean absolute error) and MAPE (mean absolute percent error) for regression and average accuracy, AUC and F1 Score for classification, as well as their standard errors. The results are summarized in table \ref{table:ENNSl1}.

We observe that our proposed algorithm enjoyed a slight performance boost via the ensemble method. Moreover, the standard errors of these results are slight greater than the standard errors reported in the last subsection, where the estimation was done assuming correct selection. The increase of standard errors is mainly due to the selection variations.

\subsection{Correlated predictors}

In this subsection, we use an example to study the numerical performance of the proposed algorithm under correlated predictors situation. We  consider two different correlations: $\rho=0.3$ and $\rho=0.7$. As a comparison, the results for $\rho=0$ is also  included. Let $u_1,...,u_n$ be i.i.d. standard normal random variables, $x_{ij}$ be i.i.d. standard normal random variables, which are independent of $u_1,...,u_n$. For $i=1,...,n$ and $j=1,...,p$, do the transformation $x_{ij}=(x_{ij}+tu_i)/\sqrt{1+t^2}$ for some $t$, then we obtain the standard normal correlated variables by 
$$cor(x_{ij},x_{ik})=\frac{t^2}{1+t^2},\ i=1,...,n;j=1,...,p$$
Taking $t=\sqrt{3/7}$ gives correlation 0.3 and taking $t=\sqrt{7/3}$ gives correlation 0.7. Then we truncate the random variables to interval $[-1, 1]$. The structure to generate response is kept the same as in the last subsection. The results of variable selection and estimation are provided in table \ref{table:correlated}.

From the table, we see that  the correlated cases perform  almost as good as there's no correlation. All models select less variables when the correlation is higher, and this is a well-known symptom of variable selection with correlated predictors. However, this does not affect the estimation step, and in some cases even this makes the estimation results better. The reason could be that we have less variables thus the model is simpler. Since the predictors are correlated, we do not lose too much information by not selecting some of them. Moreover, some results, not in the table, includes the false positive rate, where the average for ENNS is $0.05\pm 0.03$, while that of the DNP is $0.29\pm 0.14$. Therefore, ENNS includes less redundant variables in the estimation step and achieves better performance.

\begin{table}
\caption{Prediction results on the testing set using neural networks with and without $l_1$ norm regularization for $s=2, 5, 10$. RMSE is rooted mean squared error, MAE is mean absolute error, and MAPE is mean absolute percent error. Accuracy is the percentage of correct prediction, auc is area under the ROC curve, and f1 score is the inverse of inverse precision plus the inverse recall.}
    \centering
    \fbox{
    \begin{tabular}{c|ccccc}
    \hline
         Response&Metric&Method&s=2&s=5&s=10\\
         \hline
         \multirow{18}{*}{Regression}&\multirow{6}{*}{RMSE}&Neural Network&31.24(13.32)&69.46(37.40)&136.64(60.54)\\
         \hhline{~~----}
         &&Xavier initialization&18.64(11.07)&58.89(27.73)&136.58(65.57)\\
         \hhline{~~----}
         &&$l_1$ built-in&20.47(9.62)&59.37(23.61)&129.55(50.48)\\
         \hhline{~~----}
         &&\textbf{$l_1$ soft-thresholding}&\textbf{5.97(4.18)}&\textbf{45.83(33.06)}&\textbf{109.31(47.24)}\\
         \hhline{~~----}
         &&Stage-wise&10.59(11.20)&47.64(22.69)&117.65(43.96)\\
         \hhline{~~----}
         &&Bagging&25.48(10.89)&59.49(26.53)&133.45(59.72)\\
         \hhline{~-----}
         &\multirow{6}{*}{MAE}&Neural Network&16.45(10.91)&52.85(28.47)&103.76(45.99)\\
         \hhline{~~----}
         &&Xavier initialization&13.65(8.06)&45.34(22.18)&105.17(53.66)\\
         \hhline{~~----}
         &&$l_1$ built-in&15.56(7.76)&45.32(18.34)&98.54(38.36)\\
         \hhline{~~----}
         &&\textbf{$l_1$ soft-thresholding}&\textbf{4.37(3.02)}&\textbf{35.49(26.21)}&\textbf{83.85(36.45)}\\
         \hhline{~~----}
         &&Stage-wise&7.91(8.23)&38.86(20.00)&89.82(33.82)\\
         \hhline{~~----}
         &&Bagging&14.77(7.92)&43.16(20.51)&99.38(41.66)\\
         \hhline{~-----}
         &\multirow{6}{*}{MAPE}&Neural Network&0.012(0.015)&0.030(0.026)&0.033(0.026)\\
         \hhline{~~----}
         &&Xavier initialization&0.009(0.009)&0.027(0.022)&0.029(0.017)\\
         \hhline{~~----}
         &&$l_1$ built-in&0.011(0.012)&0.029(0.023)&0.032(0.021)\\
         \hhline{~~----}
         &&\textbf{$l_1$ soft-thresholding}&\textbf{0.005(0.007)}&\textbf{0.017(0.010)}&0.029(0.023)\\
         \hhline{~~----}
         &&\textbf{Stage-wise}&0.007(0.007)&0.019(0.015)&\textbf{0.027(0.016)}\\
         \hhline{~~----}
         &&Bagging&0.010(0.010)&0.026(0.024)&0.030(0.022)\\
         \hline
         \multirow{18}{*}{Classification}&\multirow{6}{*}{Accuracy}&Neural Network&0.944(0.026)&0.886(0.037)&0.841(0.041)\\
         \hhline{~~----}
         &&Xavier initialization&0.952(0.026)&0.891(0.034)&0.831(0.036)\\
         \hhline{~~----}
         &&$l_1$ built-in&0.927(0.031)&0.844(0.085)&0.752(0.110)\\
         \hhline{~~----}
         &&\textbf{$l_1$ soft-thresholding}&\textbf{0.964(0.028)}&\textbf{0.908(0.029)}&\textbf{0.855(0.031)}\\
         \hhline{~~----}
         &&Stage-wise&0.945(0.030)&0.886(0.038)&0.804(0.042)\\
         \hhline{~~----}
         &&Bagging&0.877(0.069)&0.806(0.068)&0.753(0.087)\\
         \hhline{~-----}
         &\multirow{6}{*}{AUC}&Neural Network&0.942(0.027)&0.882(0.038)&0.837(0.042)\\
         \hhline{~~----}
         &&Xavier initialization&0.951(0.027)&0.891(0.034)&0.825(0.037)\\
         \hhline{~~----}
         &&$l_1$ built-in&0.924(0.031)&0.833(0.100)&0.734(0.123)\\
         \hhline{~~----}
         &&\textbf{$l_1$ soft-thresholding}&\textbf{0.964(0.029)}&\textbf{0.905(0.029)}&\textbf{0.851(0.032)}\\
         \hhline{~~----}
         &&Stage-wise&0.943(0.031)&0.884(0.038)&0.800(0.041)\\
         \hhline{~~----}
         &&Bagging&0.877(0.065)&0.803(0.063)&0.751(0.084)\\
         \hhline{~-----}
         &\multirow{6}{*}{F1 Score}&Neural Network&0.943(0.027)&0.887(0.045)&0.841(0.049)\\
         \hhline{~~----}
         &&Xavier initialization&0.952(0.026)&0.892(0.041)&0.832(0.048)\\
         \hhline{~~----}
         &&$l_1$ built-in&0.927(0.031)&0.824(0.192)&0.732(0.200)\\
         \hhline{~~----}
         &&\textbf{$l_1$ soft-thresholding}&\textbf{0.965(0.026)}&\textbf{0.908(0.036)}&\textbf{0.857(0.033)}\\
         \hhline{~~----}
         &&Stage-wise&0.944(0.031)&0.883(0.042)&0.806(0.049)\\
         \hhline{~~----}
         &&Bagging&0.870(0.077)&0.792(0.060)&0.748(0.088)\\
    \end{tabular}}
    
    \label{table:estimationcompare}
\end{table}

\begin{table}
\caption{Model performance of the combination of ENNS algorithm and $l_1$ thresholding estimation, compared with DNP, Lasso and HSIC-Lasso for $s=2, 5, 10$ cases in both regression and classification. The average performance of 101 repetitions with their standard errors in parenthesis are presented.}
    \centering
    \fbox{
        \begin{tabular}{c|ccccc}
    \hline
         Response&Metric&Method&s=2&s=5&s=10\\
         \hline
         \multirow{12}{*}{Regression}&\multirow{4}{*}{RMSE}&\textbf{ENNS+l1}&\textbf{15.67(30.35)}&\textbf{48.14(21.16)}&\textbf{174.08(65.38)}\\
         \hhline{~~----}
         &&DNP&25.42(33.16)&62.63(29.02)&\textbf{178.91(60.15)}\\
         \hhline{~~----}
         &&Lasso&79.44(67.31)&104.19(49.38)&192.04(77.34)\\
         \hhline{~~----}
         &&HSIC-Lasso&56.32(59.41)&86.77(47.51)&188.35(56.48)\\
         \hhline{~-----}
         &\multirow{4}{*}{MAE}&\textbf{ENNS+l1}&\textbf{12.03(23.68)}&\textbf{40.12(19.95)}&\textbf{132.07(44.99)}\\
         \hhline{~~----}
         &&DNP&20.15(27.15)&47.85(22.31)&\textbf{136.06(45.95)}\\
         \hhline{~~----}
         &&Lasso&64.11(54.63)&81.97(39.76)&147.86(60.21)\\
         \hhline{~~----}
         &&HSIC-Lasso&42.89(34.66)&70.04(41.23)&144.37(48.15)\\
         \hhline{~-----}
         &\multirow{4}{*}{MAPE}&\textbf{ENNS+l1}&\textbf{0.012(0.025)}&\textbf{0.028(0.036)}&\textbf{0.041(0.037)}\\
         \hhline{~~----}
         &&DNP&0.017(0.028)&0.032(0.032)&\textbf{0.042(0.041)}\\
         \hhline{~~----}
         &&Lasso&0.042(0.029)&0.046(0.035)&0.046(0.025)\\
         \hhline{~~----}
         &&HSIC-Lasso&0.033(0.021)&0.036(0.025)&0.048(0.024)\\
         \hline
         \multirow{12}{*}{Classification}&\multirow{4}{*}{Accuracy}&\textbf{ENNS+l1}&\textbf{0.967(0.029)}&\textbf{0.848(0.025)}&\textbf{0.756(0.067)}\\
         \hhline{~~----}
         &&DNP&0.933(0.076)&0.822(0.068)&0.736(0.064)\\
         \hhline{~~----}
         &&Lasso&0.732(0.103)&0.726(0.071)&0.692(0.075)\\
         \hhline{~~----}
         &&HSIC-Lasso&0.805(0.094)&0.798(0.094)&0.706(0.081)\\
         \hhline{~-----}
         &\multirow{4}{*}{AUC}&\textbf{ENNS+l1}&\textbf{0.959(0.036)}&\textbf{0.834(0.024)}&\textbf{0.709(0.058)}\\
         \hhline{~~----}
         &&DNP&0.898(0.148)&0.780(0.100)&0.699(0.052)\\
         \hhline{~~----}
         &&Lasso&0.652(0.121)&0.640(0.102)&0.625(0.068)\\
         \hhline{~~----}
         &&HSIC-Lasso&0.774(0.125)&0.748(0.121)&0.677(0.061)\\
         \hhline{~-----}
         &\multirow{4}{*}{F1-Score}&\textbf{ENNS+l1}&\textbf{0.962(0.037)}&\textbf{0.859(0.036)}&\textbf{0.708(0.089)}\\
         \hhline{~~----}
         &&DNP&0.903(0.208)&0.761(0.199)&0.705(0.100)\\
         \hhline{~~----}
         &&Lasso&0.590(0.299)&0.604(0.250)&0.634(0.192)\\
         \hhline{~~----}
         &&HSIC-Lasso&0.744(0.206)&0.731(0.242)&0.666(0.208)\\
         \hline
    \end{tabular}}
    
    \label{table:ENNSl1}
\end{table}

\begin{table}
    \caption{Selection and estimation comparison for predictors with correlation 0, 0.3 and 0.7. The number of nonzero predictors is set to 5. For selection, the average number of correct selected variables with its standard error is given. For estimation the average RMSE or AUC with their standard errors is given. The results are averaged over 101 repetitions.}
    \centering
    \fbox{
    \begin{tabular}{c|cccc}
    \hline
         \multirow{2}{*}{Response}&\multirow{2}{*}{Model}&\multicolumn{3}{c}{selection}\\
         \hhline{~~---}
         &&$\rho=0.0$&$\rho=0.3$&$\rho=0.7$\\
         \hline
         \multirow{3}{*}{Regression}&ENNS+$l_1$&3.81(0.79)&3.27(0.75)&2.29(0.70)\\
         \hhline{~----}
         &DNP&3.48(0.96)&2.95(0.79)&2.14(0.56)\\
         \hhline{~----}
         &LASSO&3.38(0.90)&2.85(0.79)&2.11(1.12)\\
         \hline
         \multirow{3}{*}{Classification}&ENNS+$l_1$&3.66(1.05)&3.25(0.76)&2.38(0.72)\\
         \hhline{~----}
         &DNP&3.62(1.09)&3.43(0.91)&2.71(1.03)\\
         \hhline{~----}
         &LASSO&3.55(0.79)&2.90(1.31)&1.95(0.72)\\
         \hline
         \multirow{2}{*}{Response}&\multirow{2}{*}{Model}&\multicolumn{3}{c}{estimation}\\
         \hhline{~~---}
         &&$\rho=0.0$&$\rho=0.3$&$\rho=0.7$\\
         \hline
         \multirow{3}{*}{Regression}&ENNS+$l_1$&40.82(19.46)&37.17(27.29)&43.18(44.47)\\
         \hhline{~----}
         &DNP&81.43(46.00)&92.91(65.25)&101.15(90.63)\\
         \hhline{~----}
         &LASSO&131.37(74.22)&151.16(108.88)&113.30(97.54)\\
         \hline
         \multirow{3}{*}{Classification}&ENNS+$l_1$&0.856(0.040)&0.875(0.061)&0.907(0.030)\\
         \hhline{~----}
         &DNP&0.774(0.100)&0.766(0.106)&0.793(0.092)\\
         \hhline{~----}
         &LASSO&0.598(0.083)&0.634(0.117)&0.683(0.116)\\
         \hline
         
    \end{tabular}}

    \label{table:correlated}
\end{table}

\subsection{Variable selection: MRI data}
In this example, we evaluate  the proposed variable selection capability and compare it with other variable selection models in the context of a biological experiment under known ground truth. The data used in this example come from Alzheimer's Disease Neuroimaging Initiatives (ADNI), see \url{http://adni.loni.usc.edu/}. The data includes $n=265$ patients' neuroimaging results, including 164 Alzheimer's (AD) patients and 101 cognitively normal (CN) individuals. 145 regions of interest (ROIs) spanning the entire brain were calculated using Multi-Atlas ROI segmentation, and 114 ROIs were derived by combining single ROIs within a tree hierarchy to obtain volumetric measurements from larger structures. Therefore, $p=259$ ROIs were used in this example. Details of the pre-processing method can be found at \url{https://adni.bitbucket.io/reference/docs/UPENN_ROI_MARS/Multi-atlas_ROI_ADNI_Methods_mod_April2016.pdf}. Among those ROIs, biologically important features are picked, see Table \ref{table:adfeatureselection}, where red indicated most important, yellow means second important, and green means third important. The combinations and all other ROIs are not listed.

The full data set is used for variable selection, and the selection result is based on a  3-fold cross-validation sets. We run the ENNS algorithm along with the LASSO and the DNP. The selection results are presented in Table \ref{table:adfeatureselection}. Note here if a model selects a simple combination of some features, these features are also marked as selected. Moreover, Table \ref{table:adfalsepositive} shows the number of combined features selected for the models and the number of false positive selections. We observe that LASSO misses a lot of important features and selected about only one-fourth of the combined features as neural networks. This indicates that the features may have a complicated relationship with the response. ENNS performs better than the shallow DNP in terms of the metrics in table \ref{table:adfalsepositive}, where IS is a weighted average score with the weights for red, yellow and green being 3, 2 and 1, respectively; NI is the number of selected important variables; and NU is the number of selected unimportant variables. As a property of the ENNS, it selects relatively less number of  false positive variables. It's hard to track the combined features, since a lot is involved, however, the combinations represent biological intuitions. Neural network selects more combined features and perform better in this sense.

\begin{table}
    \caption{Variable selection result for the AD data. The table includes all biologically important variables with three levels: red (very important), yellow (secondly important) and green (thirdly important). The non-important variables are not included in the model. Checkmarks indicate whether the corresponding algorithm selected the variable or not.}
    \centering

    \fbox{
    \begin{tabular}{c|c|c|c|c|c|c|c|c|c|c|c|}
    \hline
Gene code&\cellcolor{yellow!}R31&\cellcolor{yellow!}R32&\cellcolor{yellow!}R36&\cellcolor{yellow!}R37&\cellcolor{green!}R38&\cellcolor{green!}R39&\cellcolor{green!}R40&\cellcolor{green!}R41&\cellcolor{red!}R47&\cellcolor{red!}R48&\cellcolor{yellow!}R49\\
Lasso&&\checkmark&\checkmark&&&&&&&\checkmark&\checkmark\\
DNP&\checkmark&\checkmark&&&&\checkmark&\checkmark&&\checkmark&\checkmark&\checkmark\\
ENNS&\checkmark&\checkmark&\checkmark&\checkmark&\checkmark&\checkmark&\checkmark&\checkmark&\checkmark&\checkmark&\checkmark\\
\hline
Gene code&\cellcolor{yellow!}R50&\cellcolor{yellow!}R51&\cellcolor{yellow!}R52&\cellcolor{green!}R55&\cellcolor{green!}R56&\cellcolor{green!}R57&\cellcolor{green!}R58&\cellcolor{green!}R59&\cellcolor{green!}R60&\cellcolor{red!}R81&\cellcolor{red!}R82\\
Lasso&\checkmark&&&\checkmark&&&&&&&\checkmark\\
DNP&&&&\checkmark&&&&&&\checkmark&\checkmark\\
ENNS&\checkmark&&&&\checkmark&&&&&\checkmark&\checkmark\\
\hline
Gene code&\cellcolor{red!}R85&\cellcolor{red!}R86&\cellcolor{red!}R87&\cellcolor{red!}R88&\cellcolor{red!}R100&\cellcolor{red!}R101&\cellcolor{green!}R102&\cellcolor{green!}R103&\cellcolor{green!}R106&\cellcolor{green!}R107&\cellcolor{red!}R116\\
Lasso&&&&&&\checkmark&&&&\checkmark&\checkmark\\
DNP&\checkmark&\checkmark&\checkmark&&&\checkmark&&\checkmark&&\checkmark&\checkmark\\
ENNS&\checkmark&\checkmark&\checkmark&\checkmark&\checkmark&\checkmark&&&\checkmark&\checkmark&\checkmark\\
\hline
Gene code&\cellcolor{red!}R117&\cellcolor{yellow!}R118&\cellcolor{yellow!}R119&\cellcolor{yellow!}R120&\cellcolor{yellow!}R121&\cellcolor{yellow!}R122&\cellcolor{yellow!}R123&\cellcolor{yellow!}R124&\cellcolor{yellow!}R125&\cellcolor{yellow!}R132&\cellcolor{yellow!}R133\\
Lasso&&&&&\checkmark&&\checkmark&\checkmark&&&\checkmark\\
DNP&\checkmark&&&\checkmark&&&&\checkmark&\checkmark&&\checkmark\\
ENNS&&&&&&&&\checkmark&\checkmark&&\checkmark\\
\hline
Gene code&\cellcolor{yellow!}R136&\cellcolor{yellow!}R137&\cellcolor{red!}R138&\cellcolor{red!}R139&\cellcolor{red!}R140&\cellcolor{red!}R141&\cellcolor{yellow!}R142&\cellcolor{yellow!}R143&\cellcolor{yellow!}R146&\cellcolor{yellow!}R147&\cellcolor{yellow!}R152\\
Lasso&\checkmark&\checkmark&\checkmark&&\checkmark&&\checkmark&&&&\\
DNP&\checkmark&\checkmark&\checkmark&\checkmark&&&\checkmark&&&&\\
ENNS&\checkmark&\checkmark&\checkmark&\checkmark&\checkmark&&\checkmark&&&&\\
\hline
Gene code&\cellcolor{yellow!}R153&\cellcolor{yellow!}R154&\cellcolor{yellow!}R155&\cellcolor{red!}R162&\cellcolor{red!}R163&\cellcolor{red!}R164&\cellcolor{red!}R165&\cellcolor{red!}R166&\cellcolor{red!}R167&\cellcolor{red!}R168&\cellcolor{red!}R169\\
Lasso&&\checkmark&\checkmark&\checkmark&&\checkmark&\checkmark&&\checkmark&&\\
DNP&&\checkmark&\checkmark&\checkmark&&&&&\checkmark&&\\
ENNS&&\checkmark&\checkmark&\checkmark&\checkmark&\checkmark&\checkmark&\checkmark&\checkmark&&\\
\hline
Gene code&\cellcolor{yellow!}R170&\cellcolor{yellow!}R171&\cellcolor{green!}R178&\cellcolor{green!}R179&\cellcolor{green!}R186&\cellcolor{green!}R187&\cellcolor{yellow!}R190&\cellcolor{yellow!}R191&\cellcolor{green!}R194&\cellcolor{green!}R195&\cellcolor{red!}R198\\
Lasso&\checkmark&\checkmark&\checkmark&\checkmark&&\checkmark&\checkmark&&&&\\
DNP&\checkmark&\checkmark&&\checkmark&&&\checkmark&\checkmark&&&\\
ENNS&\checkmark&\checkmark&&\checkmark&&&\checkmark&\checkmark&&\checkmark&\\
\hline
Gene code&\cellcolor{red!}R199&\cellcolor{yellow!}R200&\cellcolor{yellow!}R201&\cellcolor{green!}R202&\cellcolor{green!}R203&\cellcolor{green!}R204&\cellcolor{green!}R205&\cellcolor{green!}R206&\cellcolor{green!}R207&&\\
Lasso&&&&\checkmark&&\checkmark&\checkmark&&&\\
DNP&&&\checkmark&&\checkmark&\checkmark&&\checkmark&&\\
ENNS&&\checkmark&\checkmark&&\checkmark&&\checkmark&&&\\
\hline
    \end{tabular}}

\label{table:adfeatureselection}
\end{table}

\begin{table}
    \caption{Variable selection result for the AD data. The reported numbers include IS, the weighted average of selected important variables with the weights being 3, 2 and 1 for red (most important), yellow (secondly important) and green (thirdly important), respectively; NI, number of important variables selected; and NU, number of unimportant variables selected.}
    \centering
    \fbox{
    \begin{tabular}{cccc}
    \hline
         Variable selection method&IS&NI&NU\\
         \hline
         LASSO&1.094&32/86&25/59\\
         DNP&1.428&40/86&15/59\\
         ENNS&1.624&49/86&6/59\\
         \hline
    \end{tabular}}

    \label{table:adfalsepositive}
\end{table}

\section{Proof}
\label{appendix}
In this section, we will provide the proof of the theorems in section \ref{section:methodology}.
\subsection*{Proof of Proposition \ref{prop:begincompare}}
\begin{proof}
Consider independent observations $\{(\boldx_1, y_1), ..., (\boldx_n, y_n)\}$. Assume
$$x_{(j)}\sim \mathcal{N}(\boldzero, \boldI),\quad j=1,...,p$$
In the regression set up where $\boldy$ is centered, we have
$$y|x_1,...,x_p\sim\mathcal{N}(\beta_1x_1+...+\beta_px_p, \sigma^2).$$
Without loss of generality, we assume that
$$|\beta_1|\geq|\beta_2|\geq...\geq|\beta_s|$$
otherwise, we may re-arrange the order of columns of the design matrix. Furthermore, without loss of generality, we may assume all coefficients are positive, otherwise, we may multiply the corresponding column of the design matrix by $-1$. Since $s<n$, we may without loss of generality consider an orthogonal design on the matrix $\boldx_{(S)}$, which can be achieved by re-parametrization. Let $\hat{S}$ be the set of variables included in the current model. The algorithm computes
$$\boldG_{0j}=\frac{\partial}{\partial\boldW_{0j}}l(\btheta;\boldX,\boldy):=(G_{0j1},...,G_{0jK})$$
where $K$ is the size of the first hidden layer. Without loss of generality, we may consider a shallow network in this part, since there isn't any predictor $x$ involved in this section, all estimates can be treated as constants, which are universal for all $j's$. We have
$$G_{0jk}=-\frac{2}{n}\sum_{i=1}^ny_i\hat{a}_k\sigma'(\sum_{j=1}^px_{ij}\hat{\theta}_{jk}+\hat{t}_k)x_{ij},\quad k=1,...,K$$
where $\hat{a}_k,\hat{t}_k$ are estimated parameters for the initial model and $\hat{\theta}_{jk}$ is set to zero for all input variables at the very beginning. Thus we have
\begin{align*}
\|\boldG_{0j}\|_2&=\sqrt{\sum_{k=1}^K[-\frac{2}{n}\sum_{i=1}^ny_i\hat{a}_k\sigma'(\sum_{j=1}^px_{ij}\hat{\theta}_{jk}+\hat{t}_k)x_{ij}]^2}\\
&=\frac{2}{n}\sqrt{\sum_{k=1}^K\hat{a}_k^2\sigma'(\hat{t}_k)}|\boldx_{(j)}^T\boldy|
\end{align*}
Since the leading constant is independent of $j$, it's easier to consider the different part, denoted
$$c_j=|\boldx_{(j)}^T\boldy|$$
for $j\in\mathcal{C}$, where $\mathcal{C}$ is the candidate set. The first variable selected is
$$j_+=\argmax_{j\in\mathcal{C}}c_j.$$
At the very beginning, we have for $x\geq 0$ that
\begin{align}
    \prob(c_1\leq x)&=\prob\left(|\boldx_{(1)}^T\boldy|\leq x\right)\nonumber\\
    &=\prob\left(-x\leq\boldx_{(1)}^T\boldy\leq x\right)\nonumber\\
    &=\prob\left(-x\leq\beta_1+\boldx_{(1)}^T\bepsilon\leq x\right)\nonumber\\
    &=\Phi\left(\frac{x-\beta_1}{\sigma\|\boldx_{(1)}\|_2}\right)-\Phi\left(\frac{-x-\beta_1}{\sigma\|\boldx_{(1)}\|_2}\right)\nonumber\\
    &=\Phi\left(\frac{x-\beta_1}{\sigma}\right)-\Phi\left(\frac{-x-\beta_1}{\sigma}\right)
\end{align}
This result implies that greater $\beta$ leads to higher probability of large $c_1$. Then

\begin{align}
    \prob(c_1>c_2)&=\prob\left(|\boldx_{(1)}^T\boldy|>|\boldx_{(2)}^T\boldy|\right)
\end{align}
Let
\begin{equation}
    W_1 = \boldx_{(1)}^T\boldy\qquad and\qquad W_2 = \boldx_{(2)}^T\boldy
\end{equation}
which are both normally distributed. Therefore, $c_1$ and $c_2$ follow folded normal distribution
\begin{equation}
    c_1\sim FN(\beta_1, \sigma^2)\qquad and \qquad c_2\sim FN(\beta_2, \sigma^2)
\end{equation}
We can calculate
\begin{equation}
    Cov(W_1,W_2)=Cov(\beta_1+\boldx_{(1)}^T\boldy, \beta_2+\boldx_{(2)}^T\boldy)=\sigma^2\boldx_{(1)}^T\boldx_{(2)}=0
\end{equation}
Because both $W_1$ and $W_2$ are normally distributed, $W_1$ and $W_2$ are independent. Therefore, $c_1$ and $c_2$ are independent. Since both $c_1$ and $c_2$ are positive, the probability is equivalent to 
\begin{equation}
    \prob(c_1>c_2)=\prob\left(\frac{c_1}{c_2}>1\right)
\end{equation}
Let
$$c_{12}=\frac{c_1}{c_2}$$
Then we have
\begin{equation}
    c_{12}\sim RN(\beta_1, \beta_2, \sigma^2, \sigma^2)
\end{equation}
where RN stands for the ratio of folded normal distributions. By theorem 3.1 in \cite{kim2006ratio}, we have the CDF of $c_{12}$
\begin{align}
    F_{12}(x)=&2L\left(\frac{\beta_1-\beta_2x}{\sigma\sqrt{1+x^2}}, -\frac{\beta_2}{\sigma}, \frac{x}{\sqrt{1+x^2}}\right) + 2L\left(\frac{\beta_1+\beta_2x}{\sigma\sqrt{1+x^2}}, \frac{\beta_2}{\sigma}, \frac{x}{\sqrt{1+x^2}}\right) +\nonumber\\
    &\Phi\left(\frac{\beta_1-\beta_2x}{\sigma\sqrt{1+x^2}}\right) + \Phi\left(\frac{\beta_1+\beta_2x}{\sigma\sqrt{1+x^2}}\right) - 2
\end{align}
where
\begin{equation}
    L(a, b, \rho) = \prob(X_1 > a, X_2 > b)
\end{equation}
with
$$
\begin{bmatrix}
X_1\\
X_2
\end{bmatrix}
\sim\mathcal{N}
\left(\boldzero,
\begin{bmatrix}
1&\rho\\
\rho&1
\end{bmatrix}
\right)$$
Then we have
\begin{align}
    \prob(c_1<c_2)&=F_{12}(1)\nonumber\\
    &=2L\left(\frac{\beta_1-\beta_2}{\sqrt{2}\sigma}, -\frac{\beta_2}{\sigma}, \frac{1}{\sqrt{2}}\right) + 2L\left(\frac{\beta_1+\beta_2}{\sqrt{2}\sigma}, \frac{\beta_2}{\sigma}, \frac{1}{\sqrt{2}}\right) +\nonumber\\
    &\Phi\left(\frac{\beta_1-\beta_2}{\sqrt{2}\sigma}\right) + \Phi\left(\frac{\beta_1+\beta_2}{\sqrt{2}\sigma}\right) - 2
\end{align}
Release the general assumption of $\beta_j>0$ by multiply $-1$ to those which are negative, we have the absolute values back on $|\beta_j|$. This is also true for different $\beta_i$ and $\beta_j$, since we did not use the difference between the nonzero predictors and zero predictors. By the exchangeability of predictors, the result holds for all $i$ and $j$. Therefore, we have
\begin{align*}
\prob(c_j<c_k)=&2L\left(\frac{|\beta_j|-|\beta_k|}{\sqrt{2}\sigma}, -\frac{|\beta_k|}{\sigma}, \frac{1}{\sqrt{2}}\right) + 2L\left(\frac{|\beta_j|+|\beta_k|}{\sqrt{2}\sigma}, \frac{|\beta_k|}{\sigma}, \frac{1}{\sqrt{2}}\right) + \\
&\Phi\left(\frac{|\beta_j|-|\beta_k|}{\sqrt{2}\sigma}\right) + \Phi\left(\frac{|\beta_j|+|\beta_k|}{\sqrt{2}\sigma}\right) - 2
\end{align*}    
\end{proof}

\subsection*{Proof of Proposition \ref{prop:beginprobability}}
\begin{proof}
Consider independent observations $\{(\boldx_1, y_1), ..., (\boldx_n, y_n)\}$. Assume
$$x_{(j)}\sim \mathcal{N}(\boldzero, \boldI),\quad j=1,...,p$$
In the regression set up where $\boldy$ is centered, we have
$$y|x_1,...,x_p\sim\mathcal{N}(\beta_1x_1+...+\beta_px_p, \sigma^2).$$
Without loss of generality, we assume that
$$|\beta_1|\geq|\beta_2|\geq...\geq|\beta_s|$$
otherwise, we may re-arrange the order of columns of the design matrix. Furthermore, without loss of generality, we may assume all coefficients are positive, otherwise, we may multiply the corresponding column of the design matrix by $-1$. Since $s<n$, we may without loss of generality consider an orthogonal design on the matrix $\boldx_{(S)}$, which can be achieved by re-parametrization. Let $\hat{S}$ be the set of variables included in the current model. At the very beginning, we have proved in the proof of proposition \ref{prop:begincompare} that 
\begin{equation}
    c_j\sim FN(\beta_j, \sigma^2)\qquad j=1,...,p
\end{equation}
and that $c_i$ and $c_j$ are independent for $i\neq j$. Denote event
\begin{equation}
    E_k=\{c_k>\max_{i\neq k}c_i\},\ k=1,...,s
\end{equation}
It's easy to observe that $E_k$'s are mutually exclusive. Therefore, we have
\begin{align}
    &Pr(At\ least\ one\ of\ c_1,...,c_s\ is\ greater\ than\ all\ of\ c_{s+1},...,c_{p})\nonumber\\
    =&Pr\left(\bigcup_{k=1}^sE_k\right)\nonumber\\
    =&\sum_{k=1}^sPr(E_k)
\end{align}
We may calculate
$$Pr(E_k)=Pr(c_k>c_{(-k,p-1)})$$
where $c_{(-k,p-1)}$ is the largest order statistic of $c_1,...c_{(k-1)},c_{(k+1)},c_{(p)}$, which is independent of $c_k$. Let $F_{(-k,p-1)}$ and $f_{(-k,p-1)}$ be the CDF and PDF of $c_{(-k,p-1)}$, respectively, we have
$$F_{(-k,p-1)}(x)=\prod_{j\neq k}^{p}F_j(x)$$
and
$$f_{(-k,p-1)}(x)=\frac{\partial}{\partial x}\prod_{j\neq k}^{p}F_j(x)$$
where from the properties of folded normal distribution we have
$$
F_k(x)=\frac{1}{2}\left[erf\left(\frac{x+|\beta_k|}{\sqrt{2\sigma^2}}\right)+erf\left(\frac{x-|\beta_k|}{\sqrt{2\sigma^2}}\right)\right]$$
and
$$f_k(x)=\frac{\partial}{\partial x}F_k(x)=\sqrt{\frac{2}{\pi\sigma^2}}e^{-\frac{x^2+\beta_k^2}{2\sigma^2}}\cosh{\frac{\beta_k x}{\sigma^2}}
$$
Then we have
\begin{align}
&Pr(E_k)\nonumber\\
=&Pr(c_k>c_{(-k,p-1)})\nonumber\\
=&\int_0^{\infty}Pr(c_k>x)f_{(-k,p-1)}(x)dx\nonumber\\
=&\int_0^{\infty}\left[1-F_k(x)\right]\frac{\partial}{\partial x}\prod_{j\neq k}^{p}F_j(x)dx\nonumber\\
=&\left.\left[\left[1-F_k(x)\right]\prod_{j\neq k}^{p}F_j(x)\right]\right|_0^{\infty}+\int_0^{\infty}f_k(x)\prod_{j\neq k}^{p}F_j(x)dx\nonumber\\
=&\int_0^{\infty}f_k(x)\prod_{j\neq k}^{p}F_j(x)dx
\end{align}
where the second equality is by the convolution formula, the fourth equality is by integration by parts. Therefore,
\begin{align}
&Pr(At\ least\ one\ of\ c_1,...,c_m\ is\ greater\ than\ all\ of\ c_{s+1},...,c_{p})\nonumber\\
=&\sum_{k=1}^s\int_0^{\infty}f_k(x)\prod_{j\neq k}^{p}F_j(x)dx
\end{align}
Next we will show that this probability is actually a very high probability. Let
$$p_k=\int_0^{\infty}f_k(x)\prod_{j\neq k}^{p}F_j(x)dx=\E_k\left[\prod_{j\neq k}^{p}F_j(X)\right]$$
By the formulas for $F_k$ and $f_k$, we have
\begin{align}
    p_k&=\int_0^{\infty}\sqrt{\frac{2}{\pi\sigma^2}}e^{-\frac{x^2+\beta_k^2}{2\sigma^2}}\cosh{\frac{\beta_k x}{\sigma^2}}\prod_{j\neq k}\frac{1}{2}\left[erf\left(\frac{x+\beta_j}{\sqrt{2\sigma^2}}\right)+erf\left(\frac{x-\beta_j}{\sqrt{2\sigma^2}}\right)\right]dx\nonumber\\
    &=\int_0^{\infty}\sqrt{\frac{1}{2\pi\sigma^2}}\left[e^{-\frac{(x+\beta_k)^2}{2\sigma^2}}+e^{-\frac{(x-\beta_k)^2}{2\sigma^2}}\right]\prod_{j\neq k}\frac{1}{2}\left[erf\left(\frac{x+\beta_j}{\sqrt{2\sigma^2}}\right)+erf\left(\frac{x-\beta_j}{\sqrt{2\sigma^2}}\right)\right]dx\nonumber\\
\end{align}
Do change of variable $z=x/\sigma$, we have
\begin{align}
    p_k&=\int_0^{\infty}\frac{1}{\sqrt{2\pi}}\left[e^{-\frac{(z+\frac{\beta_k}{\sigma})^2}{2}}+e^{-\frac{(z-\frac{\beta_k}{\sigma})^2}{2}}\right]\prod_{j\neq k}\frac{1}{2}\left[erf\left(\frac{z+\frac{\beta_j}{\sigma}}{\sqrt{2}}\right)+erf\left(\frac{z-\frac{\beta_j}{\sigma}}{\sqrt{2}}\right)\right]dz\nonumber\\
\end{align}
Let $\Tilde{\beta}_k=\beta_k/\sigma$, without loss of generality, assume that $\infty=\beta_0\geq\beta_1\geq...\geq\beta_p\geq\beta_{p+1}=0$, we have
\begin{align}
    p_k=&\int_0^{\infty}\frac{1}{\sqrt{2\pi}}\left[e^{-\frac{(z+\Tilde{\beta}_k)^2}{2}}+e^{-\frac{(z-\Tilde{\beta}_k)^2}{2}}\right]\prod_{j\neq k}\frac{1}{2}\left[erf\left(\frac{z+\Tilde{\beta}_j}{\sqrt{2}}\right)+erf\left(\frac{z-\Tilde{\beta}_j}{\sqrt{2}}\right)\right]dz\nonumber\\
    =&\sum_{i=0}^p\int_{\beta_{i+1}}^{\beta_i}\frac{1}{\sqrt{2\pi}}\left[e^{-\frac{(z+\Tilde{\beta}_k)^2}{2}}+e^{-\frac{(z-\Tilde{\beta}_k)^2}{2}}\right]\nonumber\\
    &\prod_{j\neq k}\frac{1}{2}\left[erf\left(\frac{z+\Tilde{\beta}_j}{\sqrt{2}}\right)+\indicator_{\{j\geq i+1\}} erf\left(\frac{z-\Tilde{\beta}_j}{\sqrt{2}}\right)-\indicator_{\{j\leq i\}} erf\left(\frac{\Tilde{\beta}_j-z}{\sqrt{2}}\right)\right]dz\nonumber\\
\end{align}
By the exponential approximation of the error function, see for example \cite{tsay2013simple}, there exist $c_1$ and $c_2$ such that
$$\sup_{x>0}|erf(x)-(1-\exp[-c_1x-c_2x^2])|$$
can be arbitrarily small, where approximately $c_1\approx 1.095$ and $c_2\approx 0.7565$. Consider this approximation, we have
\begin{align}
    p_k=&\sum_{i=0}^p\int_{\beta_{i+1}}^{\beta_i}\frac{1}{\sqrt{2\pi}}\left[e^{-\frac{(z+\Tilde{\beta}_k)^2}{2}}+e^{-\frac{(z-\Tilde{\beta}_k)^2}{2}}\right]\nonumber\\
    &\prod_{j\neq k}\frac{1}{2}\left\{1+\indicator_{\{j\geq i+1\}}-\indicator_{\{j\leq i\}}-e^{\frac{c_1^2}{4c_2}}\left[e^{-\frac{c_2}{2}\left[z+\left(\tilde{\beta}_j+\frac{c_1}{\sqrt{2}c_2}\right)\right]^2}\right.\right.\nonumber\\
    &\left.\left.+\indicator_{\{j\geq i+1\}}e^{-\frac{c_2}{2}\left[z+\left(-\tilde{\beta}_j+\frac{c_1}{\sqrt{2}c_2}\right)\right]^2}-\indicator_{\{j\leq i\}}e^{-\frac{c_2}{2}\left[z-\left(\tilde{\beta}_j+\frac{c_1}{\sqrt{2}c_2}\right)\right]^2}\right]\right\}
\end{align}
Here
$$e^{\frac{c_1^2}{4c_2}}\approx 1.48>>1$$
Observe that as when $i=s$, also observe that $\tau_n\rightarrow 0$ indicates $\max_{j=s+1,...,p}\beta_j\rightarrow 0$, we have
$$\prod_{j=1, j\neq k}^s\frac{1}{2}e^{\frac{c_1^2}{4c_2}}\left[e^{-\frac{c_2}{2}\left[z-\left(\beta_j+\frac{c_1}{\sqrt{2}c_2}\right)\right]^2}-e^{-\frac{c_2}{2}\left[z+\left(\beta_j+\frac{c_1}{\sqrt{2}c_2}\right)\right]^2}\right]\rightarrow 0\ as\ s\rightarrow\infty$$
Therefore, the formula of $p_k$ can be simplified to
\begin{align}
    p_k&=o\left(\frac{1}{2^s}e^{\frac{sc_1^2}{4c_2}}\right)+\sum_{i=0}^s\int_{\beta_{i+1}}^{\beta_i}\frac{1}{\sqrt{2\pi}}\left[e^{-\frac{(z+\Tilde{\beta}_k)^2}{2}}+e^{-\frac{(z-\Tilde{\beta}_k)^2}{2}}\right]\nonumber\\
    &\prod_{j\neq k}\frac{1}{2}\left\{1+\indicator_{\{j\geq i+1\}}-\indicator_{\{j\leq i\}}-e^{\frac{c_1^2}{4c_2}}\left[e^{-\frac{c_2}{2}\left[z+\left(\tilde{\beta}_j+\frac{c_1}{\sqrt{2}c_2}\right)\right]^2}\right.\right.\nonumber\\
    &\left.\left.+\indicator_{\{j\geq i+1\}}e^{-\frac{c_2}{2}\left[z+\left(-\tilde{\beta}_j+\frac{c_1}{\sqrt{2}c_2}\right)\right]^2}-\indicator_{\{j\leq i\}}e^{-\frac{c_2}{2}\left[z-\left(\tilde{\beta}_j+\frac{c_1}{\sqrt{2}c_2}\right)\right]^2}\right]\right\}\nonumber\\
    &\leq o\left(\frac{1}{2^s}e^{\frac{sc_1^2}{4c_2}}\right)+\sum_{i=0}^s\left(\frac{1}{2}e^{\frac{c_1^2}{4c_2}}\right)^{s-i}\frac{1}{2s}\left[\Phi(\tilde{\beta}_i-\tilde{\beta}_k)-\Phi(\tilde{\beta}_{i+1}-\tilde{\beta}_k)+\Phi(\tilde{\beta}_i+\tilde{\beta}_k)-\Phi(\tilde{\beta}_{i+1}+\tilde{\beta}_k)\right]
\end{align}
where $\Phi$ is the normal CDF and the inequality is by observing
$$e^{-x^2}\leq 1$$
and the term in the bracket is less than $2$ when $j\geq i+1$. Then summing up $p_k's$ and observing the double sum is not converging to zero since it consists of a geometric component, when $\beta_{max}$ is not big enough and let $s\rightarrow\infty$, we have
\begin{align}
    &1-\sum_{k=1}^sp_k\nonumber\\
    \geq& 1-o\left(s\frac{1}{2^s}e^{\frac{sc_1^2}{4c_2}}\right)-\sum_{k=1}^s\sum_{i=0}^s\left(\frac{1}{2}e^{\frac{c_1^2}{4c_2}}\right)^{s-i}\nonumber\\
    &\frac{1}{2s}\left[\Phi(\tilde{\beta}_i-\tilde{\beta}_k)-\Phi(\tilde{\beta}_{i+1}-\tilde{\beta}_k)+\Phi(\tilde{\beta}_i+\tilde{\beta}_k)-\Phi(\tilde{\beta}_{i+1}+\tilde{\beta}_k)\right]\nonumber\\
    \geq& \sum_{i=1}^s\left(\frac{1}{2}e^{\frac{c_1^2}{4c_2}}\right)^{s-i}\sum_{k=1}^s\frac{1}{2s}(1-\Phi(\beta_{max}))-o\left(s\frac{1}{2^s}e^{\frac{sc_1^2}{4c_2}}\right)\nonumber\\
    \geq& c-o(1)
\end{align}

\end{proof}

\subsection*{Proof of Theorem \ref{thm:nofalsepositive}}
\begin{proof}
In this proof, we will show the probability that the same zero predictor appears in $k$ bagging rounds tends to zero as $k$ increases. At the first step, we have $\mathcal{C}=\{1,...,p\}$ and $\mathcal{S}=\{\}$. By proposition \ref{prop:beginprobability} we know that the probability that the first variable belongs to $\mathcal{S}_0$ converges to one under the conditions.

At the $m^{th}$ step, denote the candidate set $\mathcal{C}^m$ and the selected set $\mathcal{S}^m$. Assume that $\mathcal{S}\subset\mathcal{S}_0$. Without loss of generality, consider $\sigma^2=1$. If not, we may divide the response and coefficients by $\sigma$. Consider the first case that
$$\mathcal{C}^m\cap\mathcal{S}_0\neq\emptyset$$
Let $\mathcal{C}^m\cap\mathcal{S}_0=\{j_1,...,j_{s'}\}$. By the proof of proposition \ref{prop:beginprobability}, the probability that a zero variable is selected is at most
\begin{align*}
    \prob(\text{select\ zero\ variable})&\leq 1-\frac{\sum_{j\in\mathcal{C}^m\cap\mathcal{S}_0}e^{\beta_j}}{\sum_{j\in\mathcal{C}^m}e^{\beta_{j}}}\\
    &=1-\frac{\sum_{j\in\mathcal{C}^m\cap\mathcal{S}_0}e^{\beta_j}}{\sum_{j\in\mathcal{C}^m\cap\mathcal{S}_0}e^{\beta_j}+\sum_{j\in\mathcal{C}^m\cap\mathcal{S}_0^C}e^{\beta_j}}\\
    &=\frac{\sum_{j\in\mathcal{C}^m\cap\mathcal{S}_0^C}e^{\beta_j}}{\sum_{j\in\mathcal{C}^m\cap\mathcal{S}_0}e^{\beta_j}+\sum_{j\in\mathcal{C}^m\cap\mathcal{S}_0^C}e^{\beta_j}}\\
    &\leq \frac{(|\mathcal{C}^m|-s')e^{\tau_n}}{s'e^{\gamma_n}+(|\mathcal{C}^m|-s')}
\end{align*}
where $|\mathcal{C}^m|=O(p)$ is the cardinality of $\mathcal{C}^m$ by theorem condition, $\beta_{min}=\min_{j=1,...,s}\beta_j$, $\beta_{max}=\max_{j=1,...,s}\beta_j$ and by assumption \ref{assumption:sparsityweak}
$$\tau_n=o(\gamma_n)\leq o(\beta_{min})$$
If we have
\begin{equation}
\label{equation:falsepositivecondition1}
\frac{|\mathcal{C}^m|-s'}{s'}e^{\tau_n-\gamma_n}\rightarrow 0\ as\ n\rightarrow\infty
\end{equation}
Then we have
$$\prob(\text{select\ zero\ variable})\rightarrow 0\ as\ n\rightarrow\infty$$
In this case, the probability of false positive in the ENNS algorithm goes to zero. However, it is not always that equation \ref{equation:falsepositivecondition1} is satisfied. It happens that the signal strength of nonzero variable is not big enough. This case can be combined with the other case that
$$\mathcal{C}^m\cap\mathcal{S}_0=\emptyset$$
In this case, it is (almost) guaranteed that a zero variable will be selected in the next step. However, we will show that though a zero variable is selected, as long as the number of zero variable in $\mathcal{S}$ is not too big, which is guaranteed by the theorem condition
$$s_0\leq Cs=o(p)$$
the selected zero variables in different rounds of the bagging algorithm together with the neural network random initialization make the probability that the same zero variables appears more than the threshold number of times converges to zero. Now we have
$$\mathcal{C}^m=\{j_1,...,j_{p'}\}\subset\{s+1,...,p\}$$
Consider the scenario that all bagging rounds are independent. The residual
$$\boldy-\hat{\bmu}_{\mathcal{S}^m}$$
is not related to $x_{j_1},...,x_{j_p'}$ by assumption \ref{assumption:sparsitystrong}, where $\mathcal{S}^m$ is the selected set at the $m^{th}$ step and $\hat{\bmu}_{\mathcal{S}^m}$ is the estimated conditional expectation of $\boldy$ given $\boldx_{\mathcal{S}^m}$. Therefore, the variables $\boldx_j$, $j\in\mathcal{C}^m$ are exchangeable. We have
\begin{align}
    \prob(j\in\mathcal{S}^{m+1}\cap\mathcal{C}^m)&=\prob\left(\frac{1}{B_2}\sum_{b=1}^{B_2}\indicator_{\{c_j\geq c_{(s_0-|\mathcal{S}^m|)}\}}\geq p_s\right)\nonumber\\
    &=\sum_{k=[B_2p_s]}^{B_2}{B_2\choose k}\frac{\left(s_0-|\mathcal{S}^m|\right)^k\left(p-s_0\right)^{B_2-k}}{(p-|\mathcal{S}^m|)^{B_2}}\nonumber\\
    &\leq\exp\left(-B_2\left[(1-p_s)\log\left(\frac{(1-p_s)(p-|\mathcal{S}^m|)}{p-s_0}\right)+p_s\log\left(\frac{p_s(p-|\mathcal{S}^m|)}{s_0-|\mathcal{S}^m|}\right)\right]\right)
\end{align}
where the last inequality is by \cite{arratia1989tutorial}. Since we have
$$s_0\leq Cs=o(p)$$
then we have
$$\prob\left(j\in\mathcal{S}^{m+1}\cap\mathcal{C}^m\right)\rightarrow 0\quad as\quad s\ and\ B_2\rightarrow\infty$$
Consider the last case that there is at least one variable in $\mathcal{C}^m$ that is also in $\{1,...,s\}$ and equation \ref{equation:falsepositivecondition1} does not hold. Also consider the truth that the bagging rounds are not fully independent in practice. Consider variable $j$ and the estimator
$$\hat{s}^m_j=\indicator_{\{\|\boldG_{mj}\|_2\leq t^m\}}$$
Conditioning on the observations, there exist a fixed $t^m$ such that $\hat{s}^m_j$ indicates whether variable $j$ is not selected ($=1$) or selected ($=0$). The bagged estimator is defined as
$$\hat{s}^m_{j,B}=\E\left[\indicator_{\{\|\boldG_{mj}^*\|_2\leq t_m\}}\right]$$
where $\boldG_{mj}^*$ is $\boldG_{mj}$ evaluated on a bootstrap sample. By the uniform law of large numbers, see for example \cite{gyorfi2006distribution}, we have
\begin{equation}
\label{equation:uniflln}
\sup_{\boldx, y}\left|\frac{1}{B_2}\sum_{b=1}^{B_2}\indicator_{\{\|\boldG_{mj,b}^*\|_2\leq t_m\}}-\E\left[\indicator_{\{\|\boldG_{mj}^*\|_2\leq t_m\}}\right]\right|\rightarrow 0\quad as\ B_2\rightarrow\infty
\end{equation}
Let $\prob_n^B$ be the empirical measure of the bootstrap sample. It's easy to verify that $\hat{s}^m_j$ is a smooth function evaluated at $\prob_n^B$. By assumption \ref{assumption:uncorrelatedness}, we have independent observations. Then according to \cite{gine1990bootstrapping}, see also \cite{buhlmann2002analyzing}, there exist an increasing sequence $\{b_n\}_{n\in\mathcal{N}}$ such that
$$b_n(\|\boldG_{mj}\|_2-c_0)\rightarrow\mathcal{N}(0,\sigma_{\infty}^2)$$
for some constant $c_0<\infty$ and $\sigma_{\infty}^2<\infty$. By algebra, in the $m^{th}$ step, we have
\begin{align}
    \label{equation:gmj}
    \|\boldG_{mj}\|_2&=\sqrt{\sum_{k=1}^K[-\frac{2}{n}\sum_{i=1}^n(y_i-\hat{\mu}_i)\hat{a}_k\sigma'(\boldx_{i}^T\hat{\btheta}_{k}+\hat{t}_k)x_{ij}]^2}\nonumber\\
    &=\frac{2}{n}\sqrt{\sum_{k=1}^K\hat{a}_k^2\left[\hat{\bepsilon}^T\bSigma'(\boldx_i^T\hat{\btheta}_k+\hat{t}_k)\boldx_{(j)}\right]^2}
\end{align}
where in $\hat{\btheta}_k$, $\hat{\theta}_{jk}$ is estimated from data for $j\in\mathcal{S}^m$ and $\hat{\theta}_{jk}$ equals zero for $j\in\mathcal{C}^m$, $\hat{\mu}_i$ is the neural network estimate of $y_i$ based on $\boldx_{\mathcal{S}^m}$, $\hat{\bepsilon}$ is the prediction error based on $\boldx_{\mathcal{S}^m}$, $\bSigma'(\boldx_i^T\hat{\btheta}_k+\hat{t}_k)$ is a diagonal matrix consists of $\sigma'(\cdot)$ evaluated at $\boldx_i^T\hat{\btheta}_k+\hat{t}_k$ and $\boldx_{(j)}$ is the $j^{th}$ column of $\boldx$. We have
\begin{align}
    \hat{\bepsilon}&=\boldy-\hat{f}(\hat{\btheta}, \hat{\boldt}, \hat{\bolda}, \hat{b}, \boldx_{\mathcal{S}^m})\nonumber\\
    &=\sum_{j\in\mathcal{C}^m\cap\{1,...,s\}}\beta_j\boldx_{(j)}+\left[\sum_{j\in\mathcal{S}^m\cap\{1,...,s\}}\beta_j\boldx_{(j)}-\hat{f}(\hat{\btheta}, \hat{\boldt}, \hat{\bolda}, \hat{b}, \boldx_{\mathcal{S}^m})\right]+\bepsilon\nonumber\\
    &=\sum_{j\in\mathcal{C}^m\cap\{1,...,s\}}\beta_j\boldx_{(j)}+\bepsilon+O\left(K_n^2\sqrt{\frac{\log(nK_n)}{n}}\right)
\end{align}
Therefore, for $j\in\mathcal{C}^m\cap\{1,...,s\}$, since $\boldx_{(j)}$ is normalized and $\sigma'(\cdot)\leq 1$, by norm inequality, we have
\begin{align}
    \E\|\boldG_{mj}\|_2&\approx\E\left[\frac{2}{n}\sqrt{\sum_{k=1}^K\hat{a}_k^2\left[\sum_{j'\in\mathcal{C}^m\cap\{1,...,s\}}\beta_{j'}\boldx_{(j')}^T\bSigma'(\boldx_i^T\hat{\btheta}_k+\hat{t}_k)\boldx_{(j)}+\bepsilon^T\bSigma'(\boldx_i^T\hat{\btheta}_k+\hat{t}_k)\boldx_{(j)}\right]^2}\right]\nonumber\\
    &\geq \E\left[\frac{2}{nK}\sum_{k=1}^K|\hat{a}_k|\left|\sum_{j'\in\mathcal{C}^m\cap\{1,...,s\}}\beta_{j'}\boldx_{(j')}^T\bSigma'(\boldx_i^T\hat{\btheta}_k+\hat{t}_k)\boldx_{(j)}+\bepsilon^T\bSigma'(\boldx_i^T\hat{\btheta}_k+\hat{t}_k)\boldx_{(j)}\right|\right]\nonumber\\
    &\geq\frac{c\cdot|\mathcal{C}^m\cap\{1,...,s\}|}{nK}\gamma_n
\end{align}
For $j\in\mathcal{C}^m\cap\{s+1,...,p\}$, by Jensen's inequality, we have
\begin{align}
    \E\|\boldG_{mj}\|_2&\approx\E\left[\frac{2}{n}\sqrt{\sum_{k=1}^K\hat{a}_k^2\left[\sum_{j'\in\mathcal{C}^m\cap\{1,...,s\}}\beta_{j'}\boldx_{(j')}^T\bSigma'(\boldx_i^T\hat{\btheta}_k+\hat{t}_k)\boldx_{(j)}+\bepsilon^T\bSigma'(\boldx_i^T\hat{\btheta}_k+\hat{t}_k)\boldx_{(j)}\right]^2}\right]\nonumber\\
    &\leq\frac{2}{n}\sqrt{\E\left\{\sum_{k=1}^K\hat{a}_k^2\left[\sum_{j'\in\mathcal{C}^m\cap\{1,...,s\}}\beta_{j'}\boldx_{(j')}^T\bSigma'(\boldx_i^T\hat{\btheta}_k+\hat{t}_k)\boldx_{(j)}+\bepsilon^T\bSigma'(\boldx_i^T\hat{\btheta}_k+\hat{t}_k)\boldx_{(j)}\right]^2\right\}}\nonumber\\
    &\leq\frac{c'}{n\sqrt{K}}
\end{align}
If $\gamma_n\geq\frac{c'\sqrt{K}}{c}$, we have
$$\prob\left(\min_{j\in\mathcal{C}^m\cap\{1,...,s\}}\E\|\boldG_{mj}\|_2\geq\max_{j\in\mathcal{C}^m\cap\{s+1,...,p\}}\E\|\boldG_{mj}\|_2\right)\rightarrow 1$$
Since we have $s_0\leq Cs=o(p)$, taking $t^m$ to be the $(|\mathcal{C}^m|-|\mathcal{S}^m|)^{th}$ smallest value of $\|\boldG_{mj}\|_2,\ j\in\mathcal{C}^m$, combine this with equation \ref{equation:uniflln}, for $j\in\mathcal{C}^m\cap\{s+1,...,p\}$, we have
\begin{align}
   \frac{1}{B_2}\sum_{b=1}^{B_2}\indicator_{\{\|\boldG_{mj,b}^*\|_2\leq t_m\}}&\leq \E\left[\indicator_{\{\|\boldG_{mj}^*\|_2\leq t_m\}}\right]+\epsilon\rightarrow\Phi\left(\frac{b_n(t^m-c_0)}{\sigma_{\infty}}-Z\right)\quad as\ n\ and\ B_2\rightarrow\infty
\end{align}
where the result is by \cite{buhlmann2002analyzing}, $Z$ is standard normal random variable and $\Phi(\cdot)$ is the standard normal CDF. Observe that $b_n$ is a diverging sequence and $s_0\leq Cs=o(p)$, then we have the probability that a zero variable is selected
\begin{align}
    \label{equation:baggingconsistency}
    &\prob\left(j\in\mathcal{C}^m\cap{\mathcal{S}^{m+1}}^c\cap\{s+1,...,p\}\right)\nonumber\\
    =&\prob\left(\left.j\in\mathcal{C}^m\cap{\mathcal{S}^{m+1}}^c\cap\{s+1,...,p\}\right|\E\|\boldG_{mj}\|_2\leq t_m\right)\prob\left(\E\|\boldG_{mj}\|_2\leq t_m\right)\nonumber\\
    &+\prob\left(\left.j\in\mathcal{C}^m\cap{\mathcal{S}^{m+1}}^c\cap\{s+1,...,p\}\right|\E\|\boldG_{mj}\|_2\geq t_m\right)\prob\left(\E\|\boldG_{mj}\|_2\geq t_m\right)\nonumber\\
    \leq&\prob\left(\left.j\in\mathcal{C}^m\cap{\mathcal{S}^{m+1}}^c\cap\{s+1,...,p\}\right|\E\|\boldG_{mj}\|_2\leq t_m\right)+\prob\left(\E\|\boldG_{mj}\|_2\geq t_m\right)\nonumber\\
    \approx& 1-\Phi\left(\frac{b_n(t^m-\E\|\boldG_{mj}\|_2)}{\sigma_{\infty}}-Z\right)+\frac{s_0-|\mathcal{S}^m|-|\mathcal{C}^m\cap\{1,...,s\}|}{p-|\mathcal{S}^m|}\nonumber\\
    \rightarrow&0\quad as\ n\rightarrow\infty\ and\ B_2\rightarrow\infty
\end{align}
Therefore, the false positive rate of the ENNS algorithm goes to zero.

In the classification set up, we have similarly for equation \ref{equation:gmj} that
\begin{align}
    \|\boldG_{mj}\|_2&=\sqrt{\sum_{k=1}^K[-\frac{1}{n}\sum_{i=1}^n(y_i-\hat{\mu}_i)\hat{a}_k\sigma'(\boldx_{i}^T\hat{\btheta}_{k}+\hat{t}_k)x_{ij}]^2}\nonumber\\
    &=\frac{1}{n}\sqrt{\sum_{k=1}^K\hat{a}_k^2\left[\hat{\bepsilon}^T\bSigma'(\boldx_i^T\hat{\btheta}_k+\hat{t}_k)\boldx_{(j)}\right]^2}
\end{align}
where in $\hat{\btheta}_k$, $\hat{\theta}_{jk}$ is estimated from data for $j\in\mathcal{S}^m$ and $\hat{\theta}_{jk}$ equals zero for $j\in\mathcal{C}^m$, $\hat{\mu}_i$ is the neural network estimate of the mean of $y_i$ based on $\boldx_{\mathcal{S}^m}$, i.e.
$$\hat{\mu}_i=\sigma\left(\sum_{k=1}^K\hat{\alpha}_k\sigma(\btheta_k^T\boldx_i+t_k)+b\right),$$
$\hat{\bepsilon}$ is the prediction error based on $\boldx_{\mathcal{S}^m}$, $\bSigma'(\boldx_i^T\hat{\btheta}_k+\hat{t}_k)$ is a diagonal matrix consists of $\sigma'(\cdot)$ evaluated at $\boldx_i^T\hat{\btheta}_k+\hat{t}_k$ and $\boldx_{(j)}$ is the $j^{th}$ column of $\boldx$. The only difference between the regression set up and the classification set up is the formula for the mean. Use Taylor expansion with Lagrange remainder, we have
\begin{align*}
\sigma\left(\sum_{j=1}^s\beta_jx_j\right)=&\sigma\left(\sum_{j\in\mathcal{C}^m\cap\{1,...,s\}}\beta_jx_j\right)+\sigma'\left(\sum_{j\in\mathcal{S}^m\cap\{1,...,s\}}\beta_jx_j+\xi\sum_{j\in\{1,...,s\}/\mathcal{S}^m}\beta_jx_j\right)\\
&\sum_{j\in\{1,...,s\}/\mathcal{S}^m}\beta_jx_j
\end{align*}
for some $\xi\in(0,1)$ and
$$0<\sigma'\left(\sum_{j\in\mathcal{S}^m\cap\{1,...,s\}}\beta_jx_j+\xi\sum_{j\in\{1,...,s\}/\mathcal{S}^m}\beta_jx_j\right)<\sigma\left(\sum_{j\in\mathcal{S}^m\cap\{1,...,s\}}\beta_jx_j+\xi\sum_{j\in\{1,...,s\}/\mathcal{S}^m}\beta_jx_j\right)<1$$
Then
$$\hat{\bepsilon}=\bepsilon+\sigma'\left(\sum_{j\in\mathcal{S}^m\cap\{1,...,s\}}\beta_jx_j+\xi\sum_{j\in\{1,...,s\}/\mathcal{S}^m}\beta_jx_j\right)\sum_{j\in\{1,...,s\}/\mathcal{S}^m}\beta_jx_j+O\left(K_n^2\sqrt{\frac{\log(nK_n)}{n}}\right)$$
where $\bepsilon$ is the theoretical error of Bernoulli distribution with their means. We don't have a direct control on $\bepsilon$, but by Cauchy-Schwarz inequality we have for any $\delta>0$ that
\begin{align}
&\prob\left(\frac{1}{n}\left|\bepsilon^T\bSigma'(\boldx_i^T\hat{\btheta}_k+\hat{t}_k)\boldx_{(j)}\right|>\delta\right)\nonumber\\
\leq&\prob\left(\frac{1}{n}\max_i\bSigma'(\boldx_i^T\hat{\btheta}_k+\hat{t}_k)\left\|\bepsilon\|_2\|\boldx_{(j)}\|_2\right|>\delta\right)\nonumber\\
\leq&\prob\left(\frac{1}{n}\left\|\bepsilon\right\|_2>\delta\right)\nonumber\\
\leq&e^{-n\delta^2/8}
\end{align}
Therefore, the only difference between classification and regression is the first order approximation term
$$\sigma'\left(\sum_{j\in\mathcal{S}^m\cap\{1,...,s\}}\beta_jx_j+\xi\sum_{j\in\{1,...,s\}/\mathcal{S}^m}\beta_jx_j\right)$$
Observe that this term only depends on the true relationship and  is independent of any $j\in\mathcal{C}^m$, therefore can be treated as a constant when comparing $\|G_{mj}\|_2$. This finishes the proof for the classification case.
\end{proof}
\subsection*{Proof of Theorem \ref{thm:selectionconsistency}}
\begin{proof}
In the proof of theorem \ref{thm:nofalsepositive}, we have proved that with probability tending to $1$, the algorithm won't select any zero variables. Therefore, here it suffices to show that the model will be able to include all nonzero variables in the model. Though it looks complicated, we only need to consider the worst case:
$$\mathcal{S}^m=\{1,...,s-1\}\quad and\quad \mathcal{C}^m=\{s,s+1,...,p\}$$
and prove that variable $s$ will be selected in the next step, since variable $s$ has the smallest true coefficient $\beta_s$ among $\{1,...,s\}$ and thus all other cases have greater probability to selected a nonzero variable. Note variable $s$ will be selected
$$s\in\mathcal{S}^{m+1}\quad \Longleftrightarrow\quad \|\boldG_{ms}\|_2=\max_{j\in\mathcal{C}^m}\|\boldG_{mj}\|_2$$
Now we have
$$\|\boldG_{mj}\|_2=\frac{2}{n}\sqrt{\sum_{k=1}^K\hat{a}_k^2\left[\hat{\bepsilon}^T\bSigma'({\boldx^{s-1}_i}^T\hat{\btheta}^{s-1}_k+\hat{t}_k)\boldx_{(j)}\right]^2}$$
where ${\boldx^{s-1}}_i$ is the first $s-1^{th}$ elements in $\boldx_i$, $\hat{\btheta}^{s-1}_k$ is estimated from data as the coefficient of ${\boldx^{s-1}_i}$, $\hat{\mu}_i$ is the neural network estimate of $y_i$ based on $\boldx^{s-1}$, $\hat{\bepsilon}$ is the prediction error based on $\boldx^{s-1}$, $\bSigma'({\boldx^{s-1}_i}^T\hat{\btheta}^{s-1}_k+\hat{t}_k)$ is a diagonal matrix consists of $\sigma'(\cdot)$ evaluated at ${\boldx^{s-1}_i}^T\hat{\btheta}^{s-1}_k+\hat{t}_k$ and $\boldx_{(j)}$ is the $j^{th}$ column of $\boldx$. 

Here we need the probability that $\|\boldG_{ms}\|_2$ being the greatest among all candidates to be very big, so that it will not be missed in the ensemble filtering. For $j\in\{s+1,...,p\}$, we have
\begin{align}
&\prob\left(\|\boldG_{ms}\|_2>\|\boldG_{mj}\|_2\right)\nonumber\\
=&\prob\left(\sum_{k=1}^K\hat{\alpha}_k^2\left[\hat{\bepsilon}^T\bSigma'({\boldx^{s-1}_i}^T\hat{\btheta}^{s-1}_k+\hat{t}_k)\boldx_{(s)}\right]^2>\sum_{k=1}^K\hat{\alpha}_k^2\left[\hat{\bepsilon}^T\bSigma'({\boldx^{s-1}_i}^T\hat{\btheta}^{s-1}_k+\hat{t}_k)\boldx_{(j)}\right]^2\right)\nonumber\\
=&\prob\left(\sum_{k=1}^K\left[\left(\hat{\bepsilon}^T\hat{\alpha}_k\bSigma'({\boldx^{s-1}_i}^T\hat{\btheta}^{s-1}_k+\hat{t}_k)\boldx_{(s)}\right)^2-\left(\hat{\bepsilon}^T\hat{\alpha}_k\bSigma'({\boldx^{s-1}_i}^T\hat{\btheta}^{s-1}_k+\hat{t}_k)\boldx_{(s)}\right)^2\right]>0\right)
\end{align}
In the regression set up, observe that
$$\max_i\bSigma'_{ii}({\boldx^{s-1}_i}^T\hat{\btheta}^{s-1}_k+\hat{t}_k)=\max_i\frac{\sigma({\boldx^{s-1}_i}^T\hat{\btheta}^{s-1}_k+\hat{t}_k)}{1+\exp({\boldx^{s-1}_i}^T\hat{\btheta}^{s-1}_k+\hat{t}_k)}\leq\max_i\sigma({\boldx^{s-1}_i}^T\hat{\btheta}^{s-1}_k+\hat{t}_k)\leq 1$$
and
$$\hat{\bepsilon}=\beta_s\boldx_{(s)}+\bepsilon+O\left(K_n^2\sqrt{\frac{\log(nK_n)}{n}}\right)$$
Also by the fact that
$$A\implies B\quad\implies\quad\prob(A)\leq\prob(B)$$
we have for regression that
\begin{align}
    &\prob\left(\|\boldG_{ms}\|_2>\|\boldG_{mj}\|_2\right)\nonumber\\
    =&\prob\left(\sum_{k=1}^K\left[\left(\left(\beta_s\boldx_{(s)}+\bepsilon\right)^T\bSigma_k\boldx_{(s)}\right)^2-\left(\left(\beta_s\boldx_{(s)}+\bepsilon\right)^T\bSigma_k\boldx_{(j)}\right)^2\right]\geq O\left(KK_n^2\sqrt{\frac{\log(nK_n)}{n}}\right)\right)\nonumber\\
    =&\prob\left(\sum_{k=1}^K\left[\beta_s^2\left[\left(\boldx_{(s)}^T\bSigma_k\boldx_{(s)}\right)^2-\left(\boldx_{(s)}^T\bSigma_k\boldx_{(j)}\right)^2\right]+2\beta_s\left[\boldx_{(s)}^T\bSigma\boldx_{(s)}\bepsilon^T\bSigma(\boldx_{(s)}-\boldx_{(j)})\right]\right.\right.\nonumber\\
    &\left.\left.+\left[\left(\bepsilon^T\bSigma\boldx_{(s)}\right)^2-\left(\bepsilon^T\bSigma\boldx_{(j)}\right)^2\right]\right]\geq O\left(KK_n^2\sqrt{\frac{\log(nK_n)}{n}}\right)\right)\nonumber\\
    \geq&\prob\left(\sum_{k=1}^K\left[c'\beta_s\left[\bepsilon^T\bSigma(\boldx_{(s)}-\boldx_{(j)})\right]+\left[\left(\bepsilon^T\bSigma\boldx_{(s)}\right)^2-\left(\bepsilon^T\bSigma\boldx_{(j)}\right)^2\right]\right]\geq\right.\nonumber\\
    &\left.-cK\beta_s^2+O\left(KK_n^2\sqrt{\frac{\log(nK_n)}{n}}\right)\right)\nonumber\\
\end{align}
Observe by assumption \ref{assumption:uncorrelatedness} that $x_{(s)}$ and $x_{(j)}$ are independent and identically distributed, we have
$$\prob\left(\left(\bepsilon^T\bSigma\boldx_{(s)}\right)^2>\left(\bepsilon^T\bSigma\boldx_{(j)}\right)^2\right)=\frac{1}{2}$$
Therefore, we have
\begin{align}
    \label{equation:selectionconsistencyproofbound}
    &\prob\left(\|\boldG_{ms}\|_2>\|\boldG_{mj}\|_2\right)\nonumber\\
    \geq&\prob\left(c'\beta_s\left[\bepsilon^T\bSigma(\boldx_{(s)}-\boldx_{(j)})\right]+\left[\left(\bepsilon^T\bSigma\boldx_{(s)}\right)^2-\left(\bepsilon^T\bSigma\boldx_{(j)}\right)^2\right]\geq-c\beta_s^2+O\left(K_n^2\sqrt{\frac{\log(nK_n)}{n}}\right)\right)^K\nonumber\\
    \rightarrow&\Phi\left(\frac{c\beta_s}{\|\bSigma(\boldx_{(s)}-\boldx_{(j)})\|_2}\right)^K\geq\left(1-\delta_n\right)^{1/(p-s)}\quad as\ n\rightarrow\infty
\end{align}
under the theorem conditions for some asymptotically negligible sequence $\delta_n>0$. Then consider the bagging process, similar to \ref{equation:baggingconsistency}, according to theorem 6 in \cite{biau2008consistency}, we have
\begin{align}
    \label{equation:selectionconsistencyregressionresult}
    &\prob\left(s\notin\mathcal{C}^m\cap\mathcal{S}^{m+1}\cap{\mathcal{C}^{m+1}}^c\right)\nonumber\\
    =&\prob\left(\frac{1}{B_2}\sum_{b=1}^{B_2}\indicator_{\{\|\boldG_{ms}\|_2\neq\max_{j\in\mathcal{C}^m}\|\boldG_{mj}\|_2\}}\geq 1-p_r\right)\nonumber\\
    \leq&\frac{1}{1-p_r}\E\left[\frac{1}{B_2}\sum_{b=1}^{B_2}\indicator_{\{\|\boldG_{ms}\|_2\neq\max_{j\in\mathcal{C}^m}\|\boldG_{mj}\|_2\}}\right]\nonumber\\
    \leq&\frac{\delta_n}{1-p_r}\rightarrow 0\quad as\ n\rightarrow\infty\ and\ B_2\rightarrow\infty
\end{align}
where the first inequality is by Markov's inequality, and the second inequality is by \ref{equation:selectionconsistencyproofbound}. Therefore, the probability that variable $s$ will not enter the model in the next step tends to zero, thus with probability tending to 1, all nonzero variables are selected in the regression set up.

Consider the classification case, we have the same as in regression but
$$\hat{\bepsilon}=\bepsilon+\sigma'\left(\sum_{j\in\mathcal{S}^m\cap\{1,...,s\}}\beta_jx_j+\xi\sum_{j\in\{1,...,s\}/\mathcal{S}^m}\beta_jx_j\right)\sum_{j\in\{1,...,s\}/\mathcal{S}^m}\beta_jx_j+O\left(K_n^2\sqrt{\frac{\log(nK_n)}{n}}\right)$$
by the proof of theorem \ref{thm:nofalsepositive}, where $\bepsilon$ is the theoretical error of Bernoulli distribution with their means. Here we no longer have the normality and have an extra term $\sigma'$ which can be treated as constant in this step, but by the central limit theorem we have
$$\sqrt{n}(\boldx_{(s)}-\boldx_{(j)})^T\bSigma\bepsilon^T\Rightarrow N(0,\boldV)$$
where $\boldV$ is bounded by assumption \ref{assumption:designmatrix} and the fact that $\bSigma$ is diagonal with the largest element less than 1. Feeding this back into \ref{equation:selectionconsistencyproofbound}, we have
$$\prob(\|\boldG_{ms}\|_2>\|\boldG_{mj}\|_2)\geq(1-\delta'n)^{1/(p-s)}$$
where $\delta'_n$ is greater than $\delta_n$ up to a factor of constant but still converges to zero as $n\rightarrow\infty$, under theorem conditions. Then similar to \ref{equation:selectionconsistencyregressionresult}, we have
$$\prob\left(s\notin\mathcal{C}^m\cap\mathcal{S}^{m+1}\cap{\mathcal{C}^{m+1}}^c\right)\leq\frac{\delta_n'}{1-p_r}\rightarrow 0\quad as\ n\rightarrow\infty\ and\ B_2\rightarrow\infty$$
This finishes the proof for the classification case.
\end{proof}
\subsection*{Proof of Theorem \ref{thm:consistency}}
\begin{proof}
In this subsection, we prove the estimation and prediction of regression and classification, respectively. In the regression set up, under assumption \ref{assumption:sparsitystrong}, we have
$$\boldy=f(\boldx)+\epsilon=f(\boldx_{\mathcal{S}})+\epsilon$$
We have
\begin{align}
\label{equation:regressionconsistencydecomposition}
    &\prob\left(\E\int|f_n(\boldx_{\hat{\mathcal{S}}})-f(\boldx_{\mathcal{S}})|^2\mu(dx)\rightarrow 0\right)\nonumber\\
    =&\prob\left(\left.\E\int|f_n(\boldx_{\hat{\mathcal{S}}})-f(\boldx_{\mathcal{S}})|^2\mu(dx)\rightarrow 0\right|\hat{\mathcal{S}}=\mathcal{S}\right)\prob\left(\hat{\mathcal{S}}=\mathcal{S}\right)\nonumber\\
    &+\prob\left(\left.\E\int|f_n(\boldx_{\hat{\mathcal{S}}})-f(\boldx_{\mathcal{S}})|^2\mu(dx)\rightarrow 0\right|\hat{\mathcal{S}}\neq\mathcal{S}\right)\prob\left(\hat{\mathcal{S}}\neq\mathcal{S}\right)\nonumber\\
    \geq&\prob\left(\left.\E\int|f_n(\boldx_{\hat{\mathcal{S}}})-f(\boldx_{\mathcal{S}})|^2\mu(dx)\rightarrow 0\right|\hat{\mathcal{S}}=\mathcal{S}\right)\prob\left(\hat{\mathcal{S}}=\mathcal{S}\right)\nonumber\\
    =&\prob\left(\E\int|f_n(\boldx_{\mathcal{S}})-f(\boldx_{\mathcal{S}})|^2\mu(dx)\rightarrow 0\right)\prob\left(\hat{\mathcal{S}}=\mathcal{S}\right)
\end{align}
Observe that
$$\left|\hat{\bbeta}\right|\leq\left|\btheta\right|_1\leq K_n$$
According to \cite{gyorfi2006distribution}, when we perform a neural network estimation on the true subset of variables, we have that the total error is bounded by the approximation error, which is bounded according to \cite{fan2020universal}, plus the estimation error, which is bounded by the covering number, then by the packing number, then by the Vapnik-Chervonenkis dimension, and finally by the space dimension, i.e.
\begin{align}
&\E\int|f_n(\boldx_{\mathcal{S}})-f(\boldx_{\mathcal{S}})|^2\mu(dx)\nonumber\\
=&O\left(L\sqrt{\frac{k_n}{n-1}}\right)+\delta_n
\end{align}
where $L$ is the Lipshitz continuity coefficient, $k_n$ is the first hidden layer size, and by \cite{gyorfi2006distribution} $\delta_n$ satisfies
$$\prob\left\{\sup\delta_n>\epsilon\right\}\leq 8\left(\frac{384K_n^2(k_n+1)}{\epsilon}\right)^{(2s+5)k_n+1}e^{-n\epsilon^2/128\cdot2^4K_n^4}$$

Under theorem assumptions, the probability above is summable, thus we have the first probability in \ref{equation:regressionconsistencydecomposition} converges to 1. On the other hand, by theorem \ref{thm:selectionconsistency}, we have the second probability in \ref{equation:regressionconsistencydecomposition} converges to 1. Therefore, the result for regression set up is proved.

In the classification set up, similarly, we have
\begin{align}
\label{equation:classificationconsistencydecomposition}
    &\prob\left(R(f_{n,\hat{\mathcal{S}}})-R(f^*_{\mathcal{S}})\rightarrow 0\right)\nonumber\\
    =&\prob\left(\left.R(f_{n,\hat{\mathcal{S}}})-R(f^*_{\mathcal{S}})\rightarrow 0\right|\hat{\mathcal{S}}=\mathcal{S}\right)\prob\left(\hat{\mathcal{S}}=\mathcal{S}\right)+\prob\left(\left.R(f_{n,\hat{\mathcal{S}}})-R(f^*_{\mathcal{S}})\rightarrow 0\right|\hat{\mathcal{S}}\neq\mathcal{S}\right)\prob\left(\hat{\mathcal{S}}\neq\mathcal{S}\right)\nonumber\\
    \geq&\prob\left(\left.R(f_{n,\hat{\mathcal{S}}})-R(f^*_{\mathcal{S}})\rightarrow 0\right|\hat{\mathcal{S}}=\mathcal{S}\right)\prob\left(\hat{\mathcal{S}}=\mathcal{S}\right)\nonumber\\
    =&\prob\left(R(f_{n,\mathcal{S}})-R(f^*_{\mathcal{S}})\rightarrow 0\right)\prob\left(\hat{\mathcal{S}}=\mathcal{S}\right)
\end{align}
By \cite{devroye2013probabilistic}, we have
$$R(f_n)-R(f^*)\rightarrow 0\quad as\ n\rightarrow\infty$$
and from theorem \ref{thm:selectionconsistency}, we have the second probability in equation \ref{equation:classificationconsistencydecomposition} tends to $1$. Combine these two results, the consistency of classification case is proved.
\end{proof}
\end{appendix}

\end{document}